# Dynamic Modeling and Vibration Analysis of Large Deployable Mesh Reflectors

J. Zhang* and C. Kazoleas *

*The University of Alabama, Tuscaloosa, Alabama, 35487-0280, U.S.A.*

W. Zhu†

*University of Maryland, Baltimore County, Baltimore, Maryland, 21250, U.S.A.*

K. Zhou‡

*The Hong Kong Polytechnic University, Hong Kong, China*

S. Yuan§

*The University of Alabama, Tuscaloosa, Alabama, 35487-0280, U.S.A.*

**Large deployable mesh reflectors are essential for space applications, providing precise reflecting surfaces for high-gain antennas used in satellite communications, Earth observation, and deep-space missions. During on-orbit missions, active shape adjustment and attitude**

---

*Graduate Student, Department of Aerospace Engineering and Mechanics, Student Member of AIAA.

†Professor, Department of Mechanical Engineering.

‡Assistant Professor, Department of Civil and Environmental Engineering.

§Assistant Professor, Department of Aerospace Engineering and Mechanics, Member of AIAA (Corresponding author, sichen.yuan@ua.edu).

**control are crucial for maintaining surface accuracy and proper orientation for these reflectors, ensuring optimal performance. Preventing resonance through thorough dynamic modeling and vibration analysis is vital to avoid structural damage and ensure stability and reliability. Existing dynamic modeling approaches, such as wave and finite element methods, often fail to accurately predict dynamic responses due to the limited capability of handling three-dimensional reflectors or the oversimplification of cable members of a reflector. This paper proposes the Cartesian spatial discretization method for dynamic modeling and vibration analysis of cable-network structures in large deployable mesh reflectors. This method defines cable member positions as a summation of internal and boundary-induced terms within a global Cartesian coordinate system. Numerical simulation on a two-dimensional cable-network structure and a center-feed mesh reflector demonstrates the superiority of the proposed method over traditional approaches, highlighting its accuracy and versatility, and establishing it as a robust tool for analyzing three-dimensional complex reflector configurations.**

## Nomenclature

| | | |
|---|---|---|
| $\sigma$ | = | tension force of cable member |
| $\sigma_{\mathrm{des}}$ | = | desired tension force of cable member |
| $M_{\cos}$ | = | equilibrium matrix consisting of direction cosines |
| $\xi$ | = | dimensionless spatial variable |
| $\alpha, \beta$ | = | prescribed coefficients of a general second-order one-dimensional continuous system |
| $u$ | = | dependent variable, displacement or position of a differential element of cable member |
| $\tilde{u}$ | = | internal term of $u$ |
| $\theta$ | = | interpolation function |
| $e$ | = | boundary motion |
| $\varphi$ | = | trail function |
| $q$ | = | generalized coordinate |

$\hat{u}$ = boundary-induced term of $u$

$x, y, z$ = nodal coordinate

$\vec{R}$ = longitudinal direction of cable member

$\rho$ = density of cable member material

$E$ = Young's modulus of cable member material

$A$ = cross-sectional area of cable member

$N_l$ , $N_t$ = positive integers that control the complexity and accuracy of the CSD method

$\vec{r}$ = unit vector representing the longitudinal direction of cable member

$\vec{w}_1$ , $\vec{w}_2$ = unit vectors representing the two transverse directions of cable member

$L$ = length of cable member

$\dot{u}$ = velocity of a differential element of cable member

$dT$ = kinetic energy of a differential element of cable member

$u_x$, $u_y$, $u_z$ = $x$-, $y$-, and $z$-coordinates of $u$

$dL$ = deformed length of a differential element of cable member

$L_0$ = undeformed length of cable member

$dL_0$ = undeformed length of a differential element of cable member

$dw_c^l$ = work done by the conservative internal force on a differential element of cable member

$\varepsilon$ = strain of a differential element of cable member

$dS$ = elongation of a differential element of cable member

$dW$ = transverse displacement of a differential element of cable member

$dw_c^t$ = work done by member tension force associated with transverse displacements

$V$ = potential energy of cable member

$L_L$ = Lagrangian

$f_{nc}$ = generalized force associated with nonconservative loads

$M$ = linear mass matrix of cable member

$K$ = linear stiffness matrix of cable member

*Subscripts*

des = desired tension

cos = direction cosines

*nc* = nonconservative loads

## I. Introduction

LARGE deployable mesh reflectors (DMRs) have emerged as critical components in various space applications due to their unique ability to unfold and achieve large apertures from compact stowed configurations. Over the past few decades, there has been significant research and development dedicated to enhancing the performance and capabilities of these reflectors [1-5]. DMRs are primarily designed to create precise radio-frequency surfaces, typically spherical or parabolic [6], which are essential for high-gain antennas used in satellite communications, Earth observation, and deep-space missions. These surfaces are composed of a network of tensioned cable members forming triangular facets, facilitating the deployment process and maintaining the surface precision necessary for achieving the desired radio-frequency characteristics in the harsh environment of space. Consequently, DMRs are integral to advancing the functionality and effectiveness of spaceborne antenna systems.

During an on-orbit mission of a DMR, active shape adjustment is crucial for maintaining the precise accuracy of the reflecting surface, which directly determines the overall performance of the reflector [7]. This high surface accuracy is necessary to ensure optimal focusing and signal transmission capabilities, which are vital for the success of various communication and observational applications [8, 9]. Additionally, active attitude control is another essential aspect of DMR operation in space [10], as it maintains the correct orientation of the reflector throughout its operational life. This control ensures that the DMR remains aligned with its target, preserving the integrity and effectiveness of the reflective surface under varying environmental conditions encountered on orbit [11].

A crucial aspect of ensuring effective active shape adjustment and attitude control is the prevention of resonance, which can occur when the frequencies of actuation loads align with the natural frequencies of the DMR. Resonance leads to amplified oscillations that can cause severe structural damage or even catastrophic failure of the reflector. Therefore, it is imperative to conduct thorough dynamic modeling and vibration analysis to identify the natural frequencies of DMRs. This analysis helps in designing structures that avoid resonance with expected operational frequencies, thereby preventing excessive vibrations that could compromise the structural stability and functional reliability of the DMR.

Existing research has primarily concentrated on the supporting ring structures of DMRs to understand and enhance their dynamic behavior. Wu et al. [12] performed modal analysis and implemented active vibration control using a linear-quadratic-regulator-based approach to mitigate vibrations in a cable-net reflector. Angeletti et al. [13] developed a dynamic model of a DMR supporting ring structure composed of passive two-node beam members and active piezoelectric embedded beam members using finite element analysis (FEA). Siriguleng et al. [14] investigated the vibration modal characteristics and interactions of a reduced-scale ring-truss structure model for a large DMR, comparing experimental results with FEA simulations, and revealed significant energy transfer between low and high frequency modes. Fang et al. [15] simplified a DMR into a flexible arm-supported multi-beam structure, extending a two-dimensional global mode method to a three-dimensional version for deriving a low-dimensional dynamical model and analyzed the influence of structural parameters and dynamical responses under orbital maneuvering. Morterolle et al. [16] advanced the field by conducting modal analysis of a large space reflector design through the development of a numerical model and the construction of a reduced-scale prototype, thereby validating on-orbit behavior simulations via ground testing and accounting for gravity compensation effects. Nie et al. [17] focused on the dynamic modeling and analysis of the deployment of mesh reflector antennas considering motion feasibility. This work was later extended by He et al. [18] who examined the effects of slack cable members during the deployment process of the reflector.

The dynamic modeling and vibration analysis of cable-network structure forming the reflecting surface of a DMR has not been extensively studied in existing work. There are primarily two types of approaches for this purpose: the wave approach and the FEA approach. The wave approach was first carried out by Von Flotow [19], who applied Fourier transformation to the continuum model of a structural member, formulating it as a system of coupled partial differential equations (PDEs). This transformation converts the PDEs into a system of ordinary differential equations (ODEs) in the frequency domain, allowing the system to be interpreted in terms of two traveling waves propagating in opposite directions. Subsequent detailed studies focusing on DMRs include the work of Liu et al. [20], who used a wave-based boundary control strategy to investigate the effectiveness of controlling vibrations in a cable network antenna reflector, addressing disturbances transmitted from the boundary truss during attitude maneuvers or orbit transfers. Xu et al. [21] introduced the wave-based transfer matrix method to analyze the dynamic response of large cable-network structures by calculating the out-of-plane displacements through the multiplication of transfer matrices of periodic elements and applying all boundary conditions of the structure. Tang et al. [22] presented an analytical approach for vibration control of large DMRs, initially obtaining an exact frequency-domain solution using the dynamic stiffness method and subsequently deducing formulas for calculating local wave and power flow to extract control information from the global response solution. However, these wave approaches are limited in their applicability as they can only handle planar cable nets and are not equipped to address the rigid-body motions of cable members, thus failing to provide a comprehensive solution for the dynamic complexities of DMR structures.

The initial applications of the FEA approach to dynamic modeling of cable-network structures can be traced back to the work by Leonard and Wilfred [23], which laid the groundwork for subsequent studies by Henohold and Russell [24], and Jayaraman and Knudson [25], primarily focusing on ground infrastructure applications. The FEA method was first adapted for dynamic modeling and vibration analysis of DMRs by Shi et al. [26], and Shi and Yang [27], who developed control-oriented dynamic models to facilitate active shape adjustment of these reflectors. This approach was similarly employed in a later study, where an electromechanical coupling dynamic model of a DMR was established [28]. Wang et al. [29] delved into the resonant multi-modal dynamics of cable net reflectors subjected to harmonic loads, utilizing the extended Hamilton principle to formulate nonlinear dynamic equations, performing linear modal analysis, and employing second-order asymptotic analysis to elucidate complex nonlinear behaviors, including response curve bending, jump phenomena, and instability regions. Although straightforward, these approaches are often inadequate in predicting the dynamic responses of DMRs accurately. This shortfall is largely due to the oversimplification of cable members in the developed dynamic models, where they are typically represented as two-node bar elements with only nodal displacements. In fact, to achieve a more accurate dynamic model, the cable members of a DMR should be modeled as taut strings that account for both nodal displacements, and internal displacements in the longitudinal and transverse directions. This higher level of detail is necessary to capture the complex dynamic interactions and behaviors inherent in DMR structures.

To address the aforementioned issues in literatures, this paper presents a novel method for dynamic modeling and vibration analysis of the cable-network structures that form the reflecting surfaces of large DMRs, named the Cartesian Spatial Discretization (CSD) method. This innovative approach defines the positions of cable members of a reflector as a summation of internal and boundary-induced terms within a global Cartesian coordinate system. A nonlinear dynamic model of each cable member is derived from Lagrange's equations, resulting in a system of ODEs. For vibration analysis, this dynamic model can be linearized around an equilibrium configuration of the reflector. The dynamic model for the entire structure is then assembled by integrating the common nodal coordinates of individual cable members. Unlike traditional wave approaches that are confined to planar cable nets, the CSD method accommodates three-dimensional structures through the use of a global Cartesian coordinate system. This approach facilitates straightforward structural assembly and inherently incorporates the rigid-body motions of cable members. Another significant advantage of the CSD method is its comprehensive integration of member internal displacements within the dynamic model, thereby avoiding the oversimplification of cable members that is prevalent in conventional FEA approaches. By accounting for internal displacements, the CSD method ensures more accurate prediction of the dynamic responses of DMRs. The proposed method is versatile and applicable to both simple and complex DMR configurations. This efficiency, combined with its capacity to handle three-dimensional dynamics and rigid-body motions, makes the CSD method a robust and reliable tool for the dynamic modeling and vibration analysis of large DMRs.

## II. Problem Statement

### A. Components of a Typical DMR

The DMR considered in this study, as depicted in Fig. 1, is supported by a rigid and stable flat truss upon full deployment. The DMR consists of two nets: the front net, which serves as the reflecting surface of the reflector, and the rear net. Both nets are constructed from a mesh of flat triangular facets, with facet edges formed by elastic cable elements that are interconnected at the nodes. The nodes of the front and rear nets are linked by adjustable tension ties. During the deployment process, these initially folded nets are deployed into highly stretched elastic meshes, and the tension ties are precisely adjusted to ensure that the facets of the front net form a sphere or parabola that meets the desired surface accuracy requirement under various environmental loads, such as thermal load caused by solar radiation [11, 30]. The fully deployed surface of the DMR can be effectively modeled as a tensioned truss structure, wherein the cable elements are subjected exclusively to axial tensions. Consequently, the modeling and design of this complex structure must address both geometric and material nonlinearities to ensure accurate performance and reliability under operational conditions.

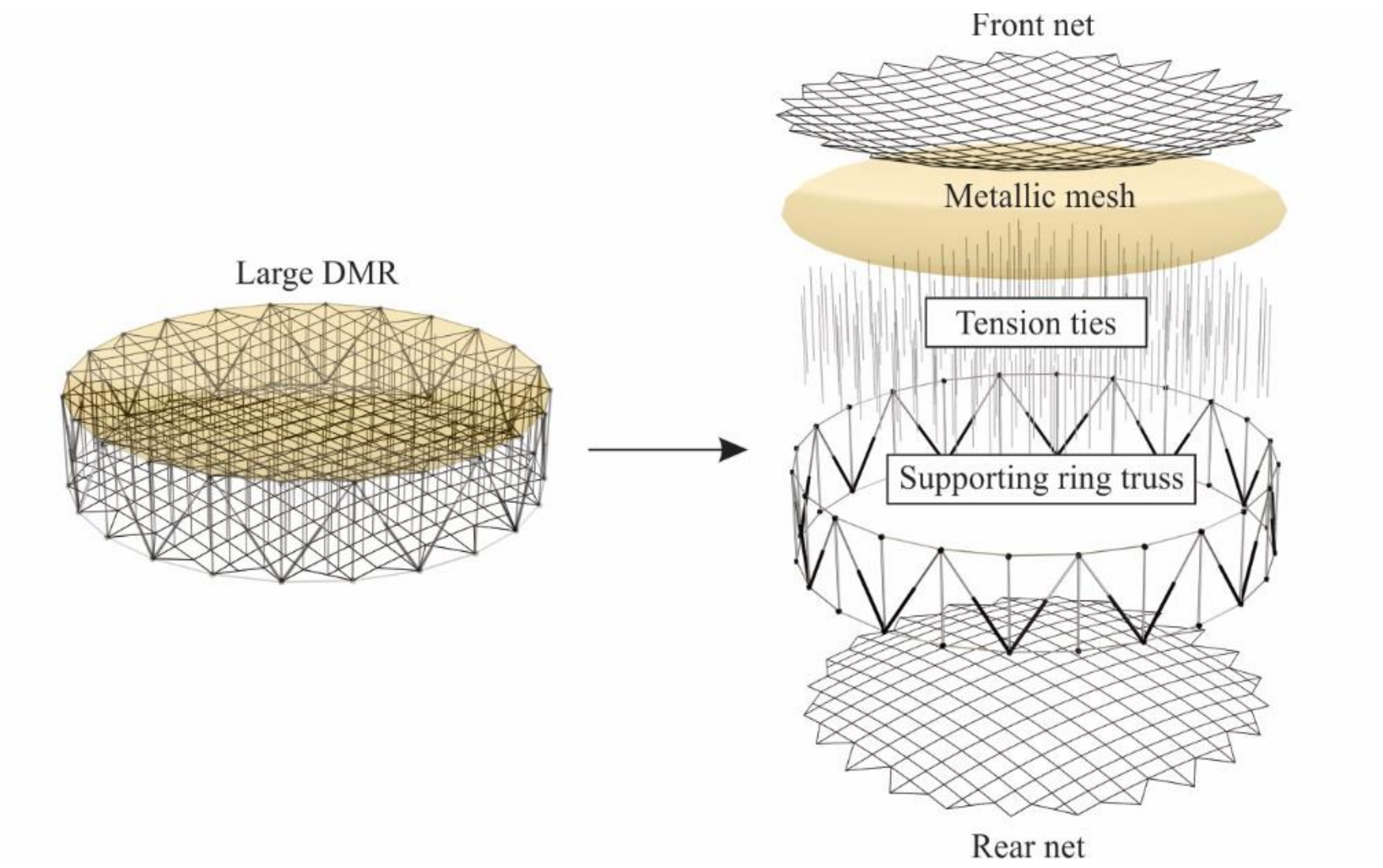

**Fig. 1 Components of a typical DMR.**

### B. Form Finding

The form finding of large DMRs involves generating precise mesh geometry meets the surface accuracy requirement and optimizing tension distribution among cables to enhance structural stiffness. The force density method [31] has traditionally been employed, simplifying nodal coordinate calculations by linearizing force equilibrium equations, while the dynamic relaxation method [32] uses fictitious masses and damping to achieve equilibrium through solving ordinary differential equations. Both methods, however, are limited by their "stress-first-and-displacement-later" approach, which restricts node placement and impedes achieving high surface accuracy. To address these limitations, researchers have modified these methods by incorporating iterative algorithms and projection techniques to refine tension distribution and mesh geometry [33-36].

Another form-finding method named fixed-nodal-position method (FNPM) [37] can maintain fixed nodal positions while producing feasible tension distributions for DMRs during the form-finding process. Unlike traditional methods that simultaneously determine nodal positions and tension distribution, this framework assigns desired nodal positions first to conform to the desired shape or surface with high accuracy, and then determines cable member tensions through a numerical optimization algorithm. Thus, the required high surface accuracy of a DMR will be achieved. The first part of the FNPM is generation of mesh geometry of a DMR by using geodesic curves to segment meshes on a reference spherical surface and then projects the meshes onto the desired working surface, ensuring superior surface accuracy [38]. The use of geodesic curves on a spherical reference surface is adopted because these curves have been proven to yield high surface accuracy in mesh reflector design, and defining geodesic curves directly on a paraboloid proves excessively challenging. This process comprises three steps: 1) Defining a spherical reference surface; 2) Generating a geodesic mesh on the reference surface; and 3) Projecting nodes from the reference surface to the desired working surface.

The nodal positions in a DMR are tightly constrained by surface accuracy requirement, leaving achieving optimal stiffness for the structure highly dependent on the second part of the FNPM, which is to determine tension distribution among the cable members by a numerical optimization algorithm. The optimization problem can be described as

$$\begin{gathered} \min \left\| \sigma - \sigma_{des} \right\|^2 \\ \text{subject to } M_{\cos}\sigma + p = 0 \quad \text{and} \quad \sigma > 0 \end{gathered} \tag{1}$$

where $\sigma$ is a vector of cable member tensions; $\sigma_{des}$ describes a desired tension distribution of cable members; and $M_{\cos}\,\sigma + p = 0$ is the force equilibrium equation of DMR in design, where $M_{\cos}$ is an equilibrium matrix consisting of direction cosines and $p$ is a vector of external loads applied by the supporting ring truss to nodes on aperture rim of the reflector. Here, the constraint $\sigma > 0$ indicates that all cable members must be tensioned. Throughout the optimization process, the nodal coordinates are unchanged. This special feature of fixed nodal positions in form-finding allows the FNPM to have guaranteed high surface accuracy in the design of DMRs.

## C. Assumptions of a DMR in Dynamic Modeling and Vibration Analysis

The objective of this research is to develop dynamic models and conduct vibration analysis of large DMRs. Properties and relevant concepts of a DMR need to be clarified for a better understanding of the proposed method. To this end, the following seven assumptions about a DMR are made:

(A1) Cable members of the reflector are connected by pin joints.

(A2) A level of tension distribution of cable members is required to stiffen the reflector and to avoid slacking of cable members.

(A3) Only tension forces are transmitted in cable members; compression, bending, and/or buckling does not occur.

(A4) The supporting ring truss is modeled as a rigid structure, assuming no deformation under dynamic simulations.

(A5) Nodal displacements of the reflector in vibration analysis are small.

(A6) Internal displacements of a cable member in both the longitudinal and transverse directions are significantly smaller than the length of the member.

(A7) Linear-elastic material behavior is assumed for all cable members.

## III. Spatial Discretization Method

Large DMRs are composed primarily of cable members, which are generally modeled as taut strings exhibiting both longitudinal and transverse displacements, characterizing them as one-dimensional continuous systems in dynamic modeling and vibration analysis. However, continuous system models governed by PDEs pose significant analytical challenges, especially for large-scale systems involving multiple components and complex coordinate transformations. This complexity is further exacerbated in large DMRs, which consist of numerous cable members arranged and connected in a three-dimensional space. The difficulty in handling PDEs for such large systems underscores the necessity for a dynamic modeling method, which facilitates efficient analysis and accurate modeling of the dynamic behavior of DMRs.

To resolve this issue, one way to consider is to transfer PDEs to ODEs which are significantly easier to solve numerically. A method based on a spatial discretization approach was developed to facilitate dynamic modeling of complex one-dimensional structural systems [39-41]. The system comprises one-dimensional distributed-parameter elements of varying lengths and lumped-parameter elements, each of which is transformed into a length-invariant component model. The governing equations for these components are derived separately and combined, with geometric matching conditions at the interfaces serving as constraints for the generalized coordinates. The Lagrangian formulation is employed to ensure that both geometric and natural matching conditions are satisfied at the component interfaces, resulting in a set of differential algebraic equations, which can be efficiently solved by a numerical solver. When all geometric matching conditions are linear, the differential algebraic equations can be transformed into a system of ODEs, which are solvable by an ODE solver.

Consider the governing equation of a second-order one-dimensional continuous systems can be written in a general form as

$$\alpha \frac{\partial^2 u(\xi,t)}{\partial t^2} + \beta \frac{\partial^2 u(\xi,t)}{\partial \xi^2} = 0, \quad \xi \in (0,1) \quad t > 0 \tag{2}$$

where $\xi$ and $t$ are the independent dimensionless spatial variable and temporal variable, respectively; $u$ is the dependent variable that denotes certain physical quantity such as displacement or position; 0 and 1 are boundary locations for the dimensionless spatial variable; and $\alpha$ and $\beta$ are prescribed coefficients. Boundary conditions of Eq. (2) are given in the general form as

$$\left.\frac{\partial^{s_1} u(\xi,t)}{\partial \xi^{s_1}}\right|_{\xi=0} = e_1(t), \quad \left.\frac{\partial^{s_2} u(\xi,t)}{\partial \xi^{s_2}}\right|_{\xi=1} = e_2(t) \tag{3}$$

where $s_1$ and $s_2$ are either 0 or 1.

Let $u(\xi, t)$ be represented in the following form as

$$u(\xi,t) = \tilde{u}(\xi,t) + \theta_1(\xi)e_1(t) + \theta_2(\xi)e_2(t) \tag{4}$$

where $\theta_i(\xi)$ ($i$ = 1, 2) are corresponding interpolation functions, $e_i(t)$ ($i$ = 1, 2) are boundary motions, and $\tilde{u}(\xi,t)$ is the internal term of function $u(\xi, t)$. The function $\tilde{u}(\xi,t)$ is defined to satisfy only simple homogeneous boundary conditions of the system:

$$\left.\frac{\partial^{s_1} \tilde{u}(\xi,t)}{\partial \xi^{s_1}}\right|_{\xi=0} = 0, \quad \left.\frac{\partial^{s_2} \tilde{u}(\xi,t)}{\partial \xi^{s_2}}\right|_{\xi=1} = 0 \tag{5}$$

By the spatial discretization method, $\tilde{u}(\xi,t)$ is expressed in an expansion form as

$$\tilde{u}(\xi,t) = \sum_{j=1}^{\infty} \varphi_j(\xi) q_j(t) \tag{6}$$

where $\varphi_j(\xi)$ ($j$ = 1, 2, …) are trial functions, which are chosen to be eigenfunctions of a simple self-adjoint system with simple homogeneous boundary conditions, and $q_j(t)$ are corresponding generalized coordinates. The functions $\theta_i(\xi)$ must be properly defined to satisfy the rule given in Eq. (7), so as to satisfy the boundary conditions in Eq. (3):

$$\begin{aligned} &\left.\frac{d^{s_1} \theta_1(\xi)}{d\xi^{s_1}}\right|_{\xi=0} = 1, \quad \left.\frac{d^{s_2} \theta_1(\xi)}{d\xi^{s_2}}\right|_{\xi=1} = 0 \\ &\left.\frac{d^{s_1} \theta_2(\xi)}{d\xi^{s_1}}\right|_{\xi=0} = 0, \quad \left.\frac{d^{s_2} \theta_2(\xi)}{d\xi^{s_2}}\right|_{\xi=1} = 1 \end{aligned} \tag{7}$$

Thus, $u(\xi, t)$ can be expressed by the sum of terms $\tilde{u}(\xi,t)$ and $\hat{u}(\xi,t)$ in the spatial discretization method:

$$u(\xi,t) = \tilde{u}(\xi,t) + \hat{u}(\xi,t) \tag{8}$$

where $\hat{u}(\xi,t)$ is the boundary-induced term of the function $u(\xi, t)$, defined as

$$\hat{u}(\xi,t) = \theta_1(\xi)e_1(t) + \theta_2(\xi)e_2(t) \tag{9}$$

According to Eq. (9), the boundary-induced term $\hat{u}(\xi,t)$ is interpolated from boundary degrees of freedom $e_i$(t) ($i$ = 1 or 2 for a second-order system), which are the dependent variable and/or its spatial derivatives at the boundaries. By substituting Eq. (9) into Eq. (7), $\hat{u}(\xi,t)$ must satisfy the boundary conditions of the original system:

$$\left.\frac{\partial^{s_1}\hat{u}(\xi,t)}{\partial\xi^{s_1}}\right|_{\xi=0}=\left.\frac{\partial^{s_1}u(\xi,t)}{\partial\xi^{s_1}}\right|_{\xi=0},\quad \left.\frac{\partial^{s_2}\hat{u}(\xi,t)}{\partial\xi^{s_2}}\right|_{\xi=1}=\left.\frac{\partial^{s_2}u(\xi,t)}{\partial\xi^{s_2}}\right|_{\xi=1} \tag{10}$$

By substituting Eqs. (6) and (9) into Eq. (8), the expression of $u(\xi,t)$ has an expansion form

$$u(\xi,t)=\sum_{j=1}^{\infty}\varphi_j(\xi)q_j(t)+\theta_1(\xi)e_1(t)+\theta_2(\xi)e_2(t) \tag{11}$$

## IV.Cartesian Spatial Discretization Method for Three-Dimensional Cable Members

Consider a cable member that connects two nodes of a DMR in a three-dimensional global Cartesian coordinate system, see Fig. 2 (a).

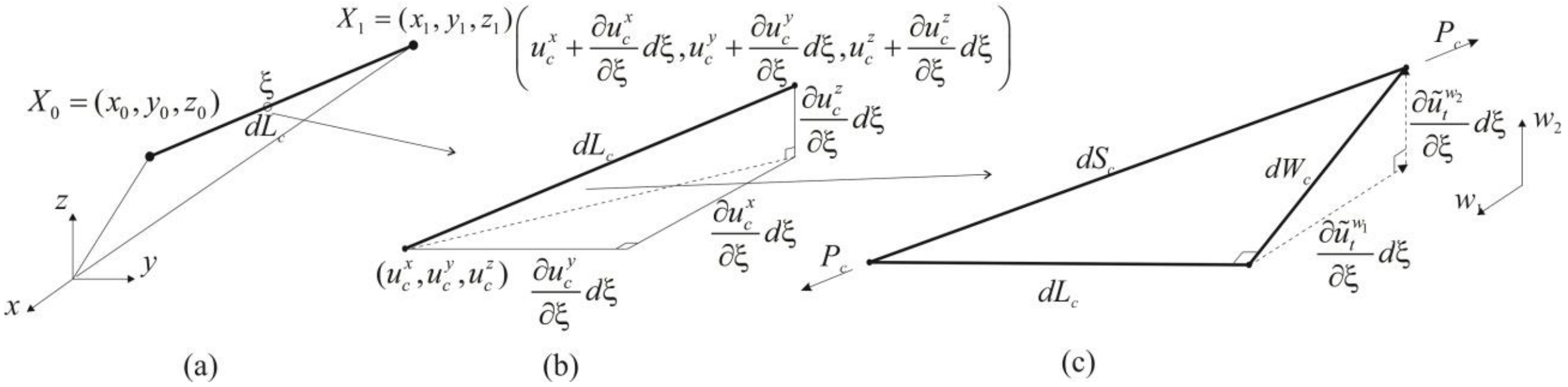

**Fig. 2 Motion of a cable member: (a) positions in the global Cartesian coordinate system, and (b) the kinetic diagram of a differential element.**

Global Cartesian coordinates of the two nodes are given as ($x_0$, $y_0$, $z_0$) and ($x_1$, $y_1$, $z_1$), respectively. The longitudinal direction of the cable member can be expressed by a position vector $\vec{R}$ :

$$\vec{R}=\begin{bmatrix}x_1-x_0\\y_1-y_0\\z_1-z_0\end{bmatrix} \tag{12}$$

An independent natural spatial variable $\xi\in[0,1]$ is used to describe an internal position of the cable member. Note that $u$ was defined by the spatial discretization method in Section III as a displacement, not a position, of a cable member, due to an assumption that the cable member is at an equilibrium state with zero displacement. However, according to the form finding process introduced in Section II, an initial equilibrium configuration of a DMR structure is determined by assignment of a set of nodal coordinates. Thus, it is convenient to define $u$ in Eq. (11) as a position of the cable member for dynamic modeling of a DMR in the three-dimensional space.

A cable member of a DMR is modeled as a taut string with longitudinal and transverse displacements. For the displacement in the longitudinal direction, the governing equation can be presented using the general form given in Eq. (2) by letting $\alpha = \rho$ and $\beta = -EA$, where $\rho$ is the linear density of the cable member material, given in mass per unit length; $E$ and $A$ are the Young's modulus and cross-sectional area of the cable member, assumed constant by assumption (A7). In dynamic modeling of a DMR, fixed-fixed boundary conditions are assumed for the cable members. Thus, trial functions in Eq. (11) is selected as $\varphi_i(\xi) = \sin(i\pi\xi)$; boundary conditions are selected to be the prescribed longitudinal displacements at the two ends: $e_1(t) = u(0,t),\quad e_2(t) = u(1,t)$; and for simplicity, the linear functions are used for the interpolation functions, given as $\theta_1(\xi) = 1 - \xi$, $\theta_2(\xi) = \xi$. For the displacement in the transverse direction, the governing equation can also be presented using the general form given in Eq. (2) by letting $\alpha = \rho$ and $\beta = -\sigma$, where $\sigma$ is the tension of the cable member. In the global Cartesian coordinate system, transverse displacements at the two ends of a cable member are considered as a combination of its rigid-body motion and longitudinal displacement in the three-dimensional space. Therefore, boundary motions in the transverse direction are assumed in a homogeneous form: $e_1(t) = 0,\quad e_2(t) = 0$ and trial functions is selected as $\varphi_j(\xi) = \sin(j\pi\xi)$.

For dynamic modeling of a cable member in global Cartesian coordinate system, Eq. (11) yields the following form:

$$u(\xi,t) = \sum_{i=1}^{N_l} \sin(i\pi\xi) q_l^i \vec{r} + \sum_{j=1}^{N_t} \sin(j\pi\xi) q_{t_1}^j \vec{w}_1 + \sum_{j=1}^{N_t} \sin(j\pi\xi) q_{t_2}^j \vec{w}_2 + (1-\xi)\begin{bmatrix} x_0 \\ y_0 \\ z_0 \end{bmatrix} + \xi \begin{bmatrix} x_1 \\ y_1 \\ z_1 \end{bmatrix} \tag{13}$$

where $N_l$ and $N_t$ are positive integers that control the complexity and accuracy of the method. The generalized coordinates $q_l$, $q_{t_1}$ and $q_{t_2}$ are employed to describe the internal displacements of the cable member in the longitudinal and transverse directions, respectively. Specifically, $q_l = 0$ indicates the absence of internal displacement in the longitudinal direction, and $(q_{t_1}, q_{t_2}) = 0$ signifies that there are no internal displacements in the two transverse directions. It is noteworthy that the transverse displacements at the two boundary nodes are maintained at zero at all times. This is due to the fact that the nodal displacement is modeled as a combination of member rigid-body motion and longitudinal displacement in the three-dimensional space using the CSD method. Consequently, if the two boundary nodes are fixed, the cable member reaches equilibrium when the generalized coordinates $q_l$, $q_{t_1}$ and $q_{t_2}$ are zero. Unit vectors $\vec{r}$ , $\vec{w}_1$ and $\vec{w}_2$ in Eqs. (13) are used represent the longitudinal direction and two transverse directions of the cable member, respectively:

$$\vec{r} = \frac{\vec{R}}{L}, \quad \vec{w}_1 = \frac{\vec{W}_1}{L_1}, \quad \vec{w}_2 = \frac{\vec{W}_2}{L_2} \tag{14}$$

where the unit vectors $\vec{w}_1$ and $\vec{w}_2$ are perpendicular to the longitudinal direction $\vec{r}$ of the cable member, and are perpendicular to each other; $L$ is the magnitude of the vector $\vec{R}$ , representing the length of the cable member subject to the longitudinal displacements:

$$L = \sqrt{(x_1 - x_0)^2 + (y_1 - y_0)^2 + (z_1 - z_0)^2} \tag{15}$$

$L_1$ and $L_2$ are magnitudes of the vectors $\vec{W}_1$ and $\vec{W}_2$ ; $\vec{W}_1$ can be defined as one of the three possible forms:

$$\vec{W}_1 = \begin{bmatrix} y_0 - y_1 \\ x_1 - x_0 \\ 0 \end{bmatrix}, \quad \vec{W}_1 = \begin{bmatrix} z_0 - z_1 \\ 0 \\ x_1 - x_0 \end{bmatrix}, \quad \vec{W}_1 = \begin{bmatrix} 0 \\ z_0 - z_1 \\ y_1 - y_0 \end{bmatrix} \tag{16}$$

and $\vec{W}_2$ is obtained by

$$\vec{W}_2 = \vec{R} \times \vec{W}_1 \tag{17}$$

The velocity $\dot{u}$ of a differential element at $\xi$ on the cable member can be obtained by taking the time derivative of $u$ in Eq. (13). Since the internal displacement described by the generalized coordinates is usually significantly smaller than the deformed length of the cable member, it can be assumed that

$$\frac{q_l}{L} \approx 0, \quad \frac{q_{t_1}}{L_1} \approx 0, \quad \frac{q_{t_2}}{L_2} \approx 0 \tag{18}$$

Therefore, the terms associated with $q_l / L$ , $q_{t_1} / L_1$ and $q_{t_2} / L_2$ vanish and velocity $\dot{u}$ becomes:

$$\dot{u}(\xi,t) = \sum_{i=1}^{N_l} \left[ \dot{q}_l^i \sin(i\pi\xi)\vec{r} \right] + \sum_{j=1}^{N_t} \left[ \dot{q}_{t_1}^j \sin(j\pi\xi)\vec{w}_1 + \dot{q}_{t_2}^j \sin(j\pi\xi)\vec{w}_2 \right] + (1-\xi) \begin{bmatrix} \dot{x}_0 \\ \dot{y}_0 \\ \dot{z}_0 \end{bmatrix} + \xi \begin{bmatrix} \dot{x}_1 \\ \dot{y}_1 \\ \dot{z}_1 \end{bmatrix} \tag{19}$$

The kinetic energy of the differential element of the cable member is

$$dT = \frac{1}{2} m \|\dot{u}\|^2 \, d\xi \tag{20}$$

where $m$ is mass of the cable member. Thus, the kinetic energy of the entire cable member can be obtained by integrating $dT$ with respect to $\xi$ in the domain [0,1]:

$$T = \int_{\xi=0}^{\xi=1} dT \tag{21}$$

A differential element of a cable member that starts at the position $\xi$ and ends at the position $\xi + d\xi$ is shown in Figs. 2 (b) and (c). The displacement of the differential element is composed of longitudinal and transverse ones. The global Cartesian coordinates of the starting and ending points of the differential element are given as ($u_x$, $u_y$, $u_z$) and $(u_x + \frac{\partial u_x}{\partial \xi} d\xi, u_y + \frac{\partial u_y}{\partial \xi} d\xi, u_z + \frac{\partial u_z}{\partial \xi} d\xi)$, respectively, where $u_x$, $u_y$, $u_z$ are the $x$-, $y$-, and $z$-coordinates of $u$. First, assume the cable member is only subjected to longitudinal displacements. An explicit expression of the deformed length $dL$ of the differential element of the cable member can be obtained by the corresponding geometry information in Fig. 2 (b):

$$dL = \left[ L + \sum_{i=1}^{N_l} q_l^i i\pi \cos(i\pi\xi) \right] d\xi \tag{22}$$

Let the undeformed length of the cable member be $L_0$, and the undeformed length of the differential element be $dL_0 = L_0 d\xi$. Then the strain $\varepsilon$ of the differential element of the member is $(dL - dL_0) / dL_0$. The work done by the conservative internal force on the differential element is given as

$$dw_c^l = \frac{1}{2} EA\varepsilon (dL - dL_0) \tag{23}$$

After the differential element elongates to the length $dL$, due to a longitudinal displacement (see Fig. 2 (b)), a transverse displacement is imposed to the same element (see Fig. 2 (c)). Under the transverse displacement, the differential element elongates to the length $dS$. According to the geometry information in Fig. 2 (c), $dS$, $dL$ and $dW$ have the following relationship as they form a right triangle:

$$dS = \sqrt{dL^2 + dW^2} \tag{24}$$

where the transverse displacement $dW$ is expressed as

$$dW = \sqrt{\left(\partial u_{w_1}^t / \partial \xi\right)^2 + \left(\partial u_{w_2}^t / \partial \xi\right)^2} \tag{25}$$

in which

$$u_{w_1}^t = \sum_{j=1}^{N_t} q_{t_1}^j \sin(j\pi\xi), \quad u_{w_2}^t = \sum_{j=1}^{N_t} q_{t_2}^j \sin(j\pi\xi) \tag{26}$$

Since the transverse displacement of the cable member usually causes little additional change in the member length, the cable tension remains the same after the transverse displacement is imposed on the differential element. Thus, the work done by the tension on the differential element of the cable member associated with the longitudinal and transverse displacements are independent of each other. The work done by the member tension force associated with the transverse displacements is

$$dw_c^t = \sigma(dS - dL) \tag{27}$$

The potential energy of the cable member is obtained by integrating $(dw_c^l + dw_c^t)$ with respect to $\xi$ in the domain [0,1]:

$$V = \int_{\xi=0}^{\xi=1} (dw_c^l + dw_c^t) \tag{28}$$

Finally, let the Lagrangian be $L_L = T - V$ ; the nonlinear equation of motion of a cable member of a DMR can be obtained by the Lagrange's Equations

$$\frac{d}{dt}\left(\frac{\partial L_L}{\partial \dot{q}}\right) - \frac{\partial L_L}{\partial q} = f_{nc} \tag{29}$$

where and $q$ are generalized coordinates defined as

$$q = \begin{bmatrix} x_0 & y_0 & z_0 & x_1 & y_1 & z_1 & q_l^1 & \cdots & q_l^{N_l} & q_{t_1}^1 & q_{t_2}^1 & \cdots & q_{t_1}^{N_t} & q_{t_2}^{N_t} \end{bmatrix} \tag{30}$$

and $f_{nc}$ is the generalized force vector associated with nonconservative loads obtained by the virtual work expression.

The nonlinear equations of motion of cable members can be linearized at an equilibrium configuration of a DMR for its vibration analysis. The linearized equation of motion is given as

$$M\ddot{q} + Kq = f_{nc} \tag{31}$$

where $M$ and $K$ are the linear mass matrix and stiffness matrix of the cable members, respectively. A dynamic model of the whole DMR can be assembled in a straight-forward way by using common nodal coordinates of structural members in Eq. (31).

## V.Numerical Examples

To demonstrate the efficacy of the CSD method in dynamic modeling and vibration analysis of large DMRs, two illustrative examples are studied. The first example involves a planar cable-network structure that serves to illustrate the capability if the CSD method and process in a controlled, simplified context. The second example focuses on a center-feed DMR, which has 101 nodes on its reflecting surface. This example aims to validate the applicability of the CSD method to real-world DMR configurations and to highlight its effectiveness in handling the complex dynamics of such structures. To evaluate the accuracy of the proposed method, the results are compared with those obtained using the FEA method in Ref. [27], which did not account for internal displacements of cable members in its dynamic modeling. This comparison underscores the advantages of incorporating internal displacements in the CSD method for precise dynamic response prediction.

### A. A planar cable-network structure

To illustrate the concepts of the CSD method, the dynamic modeling and vibration analysis of a planar cable-network structure, as depicted in Fig. 3 is considered. This structure consists of two free nodes, seven fixed nodes, and nine cable members. The structure is supported by fixed nodes three through nine. Materials of cable members are assumed to be steel. According to assumption (A7), all cable members are linear elastic, each with a Young's modulus $E$ of 200 $GPa$, a uniform cross-sectional area of 3.14 $mm^2$, and a material density of 7850 $kg/m^3$. In the form-finding process to determine the initial equilibrium configuration of the structure as introduced in Section II Part B, it is assumed that no external loads are applied. The form-finding results, including the nodal positions and member tension distribution are given in Table 1. The dynamic modeling of a planar cable member by the CSD method is given in Appendix.

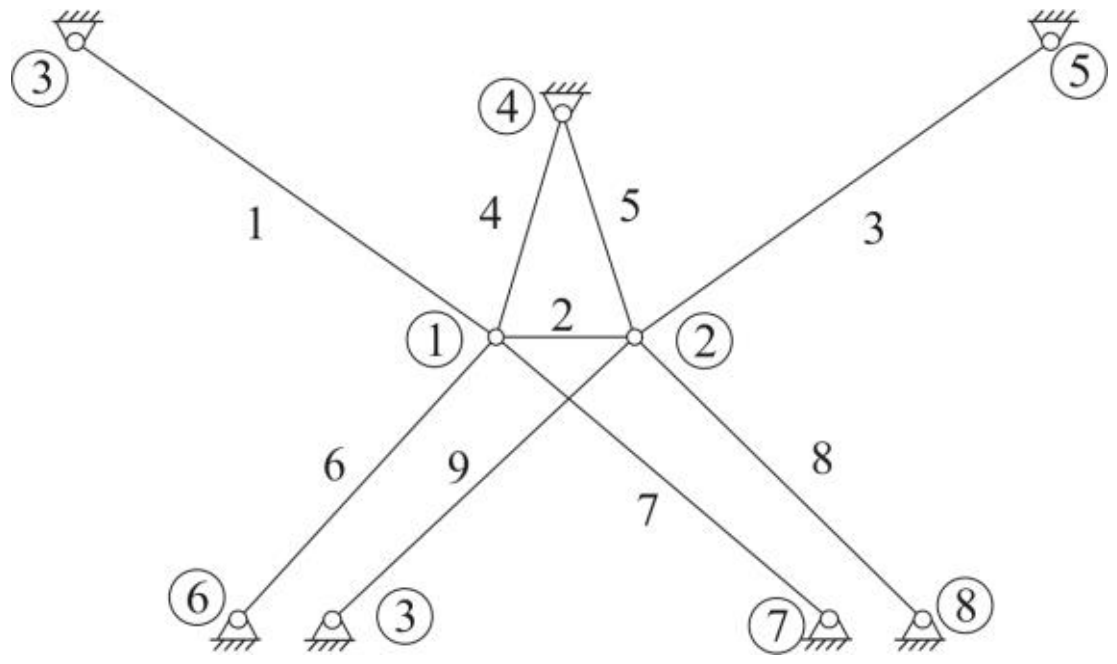

**Fig. 3 A planar cable-network structure two free nodes, seven fixed nodes, and nine cable members.**

**Table 1 Nodal coordinates and member tensions of the initial equilibrium configuration of the planar cable-network structure**

| Nodal index | *x*-coordinate (meter) | *y*-coordinate (meter) | Member Index | Tension (Newton) |
|---|---|---|---|---|
| 1 | -0.1228 | -0.1964 | 1 | 626.15 |
| 2 | 0.1228 | -0.1964 | 2 | 368.80 |
| 3 | -1.5000 | 1.0000 | 3 | 626.15 |
| 4 | 0 | 0.5000 | 4 | 598.45 |
| 5 | 1.5000 | 1.0000 | 5 | 598.45 |
| 6 | -1.1228 | -1.1964 | 6 | 707.10 |
| 7 | 0.8772 | -1.1964 | 7 | 707.10 |
| 8 | 1.1228 | -1.1964 | 8 | 707.10 |
| 9 | -0.8772 | -1.1964 | 9 | 707.10 |

The modal analysis of the structure is conducted using both the CSD method and the FEA method. The mode shapes of the structure, obtained by the CSD method with parameters $N_l = 0$ and $N_t = 1$, are illustrated in Figs. 4 and 5. The dynamic model, comprising 13 degrees of freedom—four from nodal displacements and nine from the transverse displacements of the cable members—exhibits 13 distinct mode shapes. Conversely, the mode shapes derived using the FEA method are displayed in Fig. 6. The dynamic model, which considers only two free nodes, each with two degrees of freedom, results in a total of four mode shapes. Upon analyzing the results, it is evident that the first nine mode shapes obtained via the CSD method are local modes, characterized by internal member displacements without significant nodal displacements. In contrast, modes 10 through 13 reveal that nodal displacements are coupled with member internal displacements, underscoring the necessity of incorporating cable member internal displacements in dynamic modeling. This coupling highlights the complexity and the interdependent nature of the structural dynamics, which is not captured by the FEA method. The FEA method, limited to revealing mode shapes associated solely with nodal displacements, omits a substantial portion of the dynamic characteristics of the structure.

The natural frequencies of a planar cable-network structure are also analyzed using both the FEA and CSD methods. A comparative analysis of the first four natural frequencies, obtained through these methods, is presented in

Table 2. It is observed that the natural frequencies derived from the CSD method are lower than those obtained via the FEA method. This discrepancy arises due to the nature of the local modes; the CSD method captures the internal displacements of the cable members, which inherently possess lower natural frequencies compared to the modes associated with nodal displacements. Furthermore, the analysis extends to the fifth through the 13th natural frequencies obtained by the CSD method, as shown in Table 3. These additional frequencies further illustrate the complex dynamic behavior of the structure, underscoring the significance of considering member internal displacements in the dynamic modeling of cable-network structures. The inclusion of these internal displacements reveals a more comprehensive and nuanced understanding of the dynamic characteristics of the structure, which the FEA method, focused predominantly on nodal displacements, fails to capture entirely.

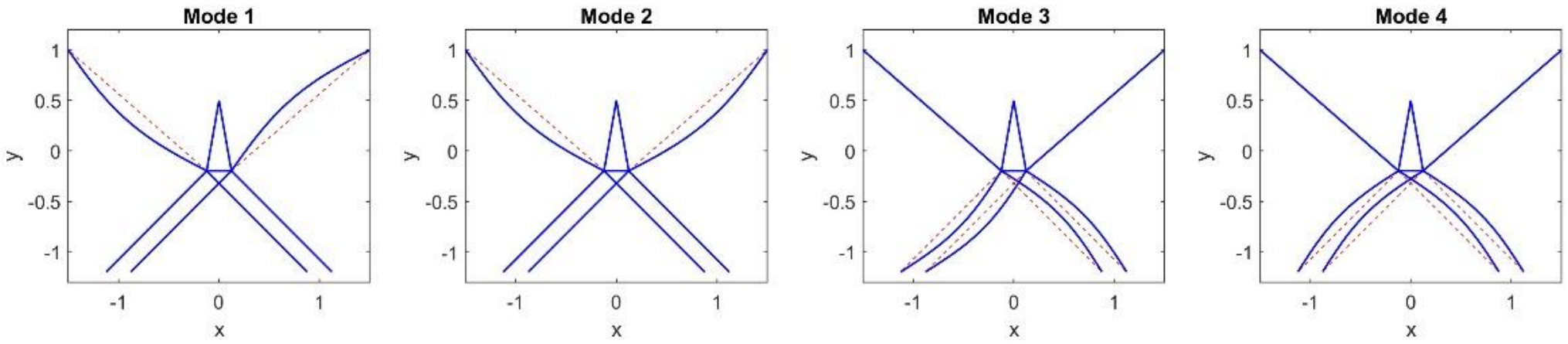

**Fig. 4  The first four mode shapes of the planar cable-network structure obtained by the CSD method.**

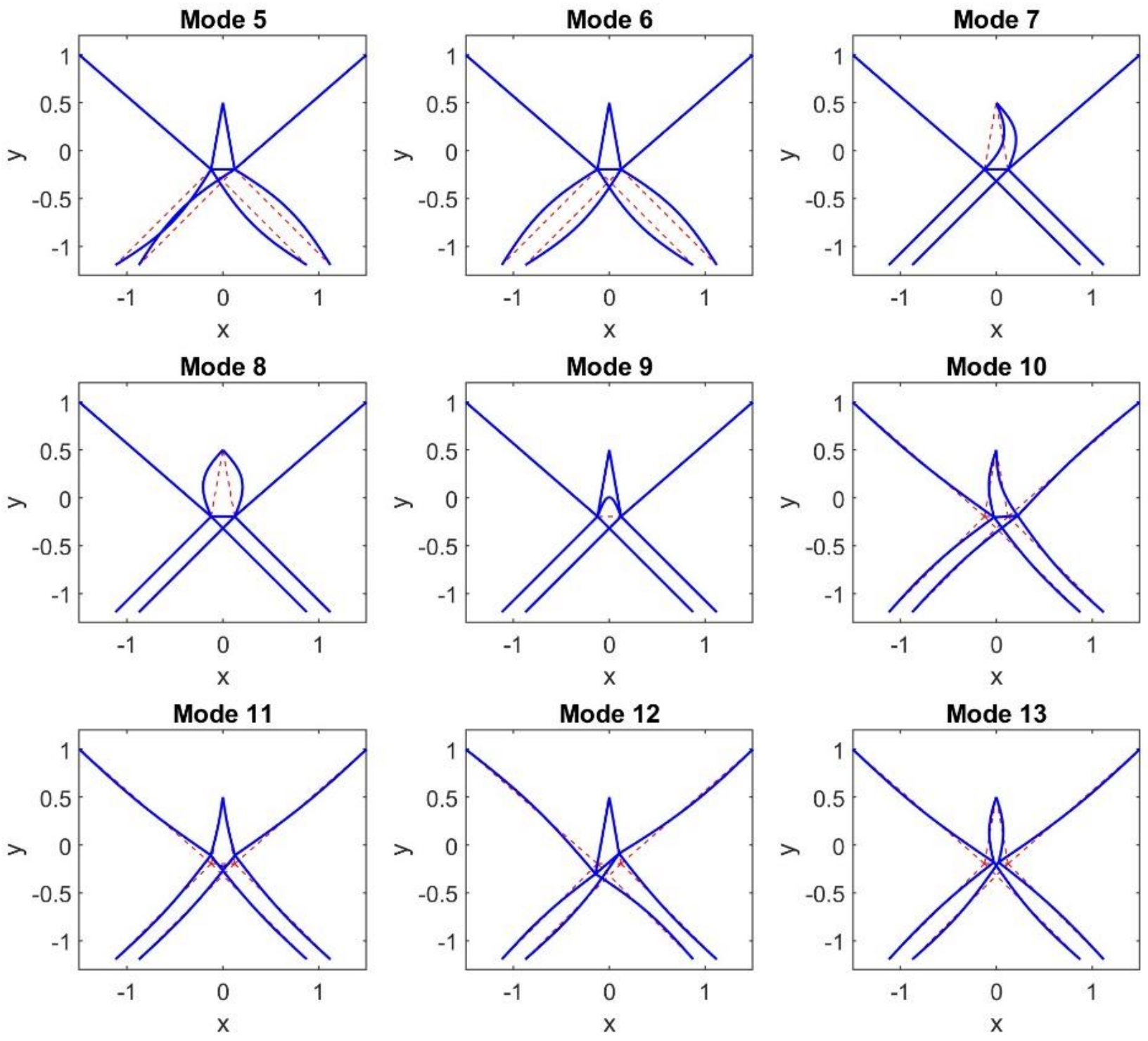

**Fig. 5 The fifth to the 13th mode shapes of the planar cable-network structure obtained by the CSD method.**

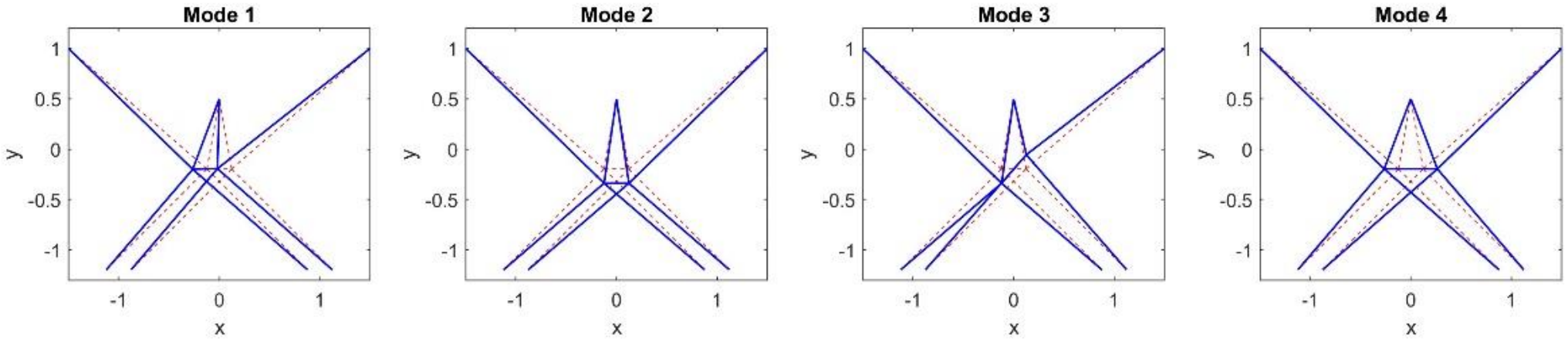

**Fig. 6 The first four mode shapes of the planar cable-network structure obtained by the FEA method.**

**Table 2 The first four natural frequencies in Hz of the planar cable-network structure obtained by the FEA method and the CSD method**

| | $f_1$ | $f_2$ | $f_3$ | $f_4$ |
|---|---|---|---|---|
| FEA method | 600.26 | 888.53 | 906.24 | 1804.73 |
| CSD method | 43.68 | 43.69 | 59.88 | 59.88 |

**Table 3 The fifth to 13th natural frequencies in Hz of the planar cable-network structure obtained by the CSD method**

| $f_5$ | $f_6$ | $f_7$ | $f_8$ | $f_9$ |
|---|---|---|---|---|
| 59.88 | 59.90 | 110.06 | 110.18 | 248.49 |
| $f_{10}$ | $f_{11}$ | $f_{11}$ | $f_{11}$ | |
| 719.89 | 1070.61 | 1075.61 | 2210.23 | |

The frequency response analysis of the structure was conducted by applying a point-wise sinusoidal force $F_f = F_0\sin(2\pi ft)$ at node two in the $x$-direction, where $F_0 = 500N$. The frequency response of node one in the $y$-direction is predicted using both the FEA and CSD methods. As illustrated in Fig. 7 (a), the results from the FEA method and the CSD method for $N_l$ = 1-3 show good agreement under 500 Hz. This concordance occurs because the natural frequencies associated with the longitudinal displacements of cable members, when internal terms are considered, are above 800 Hz. However, Fig. 7 (b) demonstrates a notable discrepancy between the FEA method and the CSD method when internal terms of cable member transverse displacements are incorporated ($N_t > 0$) in the CSD method. This mismatch under 500 Hz is attributed to the lower natural frequencies associated with the transverse displacements of the cable members, which fall below 60 Hz.

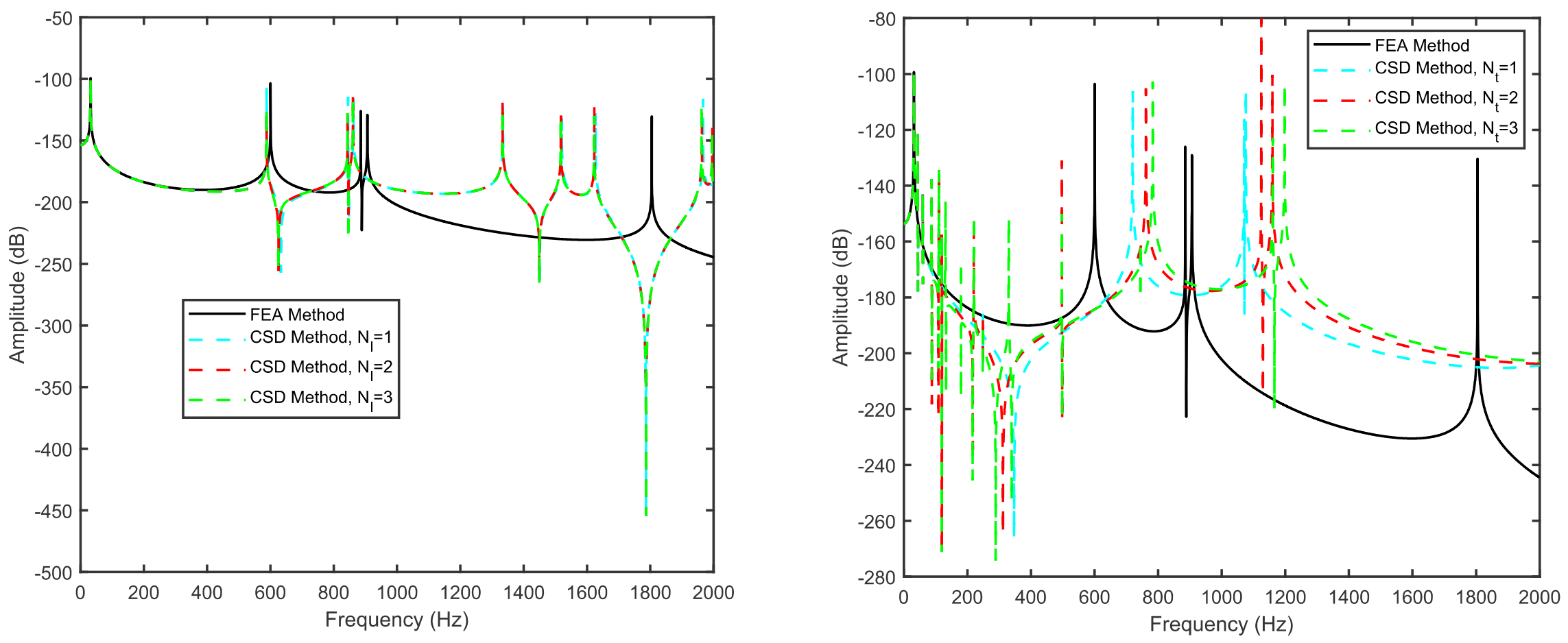

(a) (b)

**Fig. 7 Frequency responses at node one of the planar cable-network structure at 0-2000 Hz obtained by the FEA method and the CSD method for: (a) $N_l$ = 1-3 and $N_t$ = 0; and (b) $N_l$ = 0, $N_t$ = 1-3.**

The transient response analysis of the structure is conducted by applying a point-wise sinusoidal force $F_f = F_0\sin(2\pi ft)$ at node two in the *x*-direction, where $F_0 = 1000N$. The transient responses of node one in the *x*- and *y*-directions are plotted in Figs. 9-11 at excitation frequencies of 50 Hz and 650 Hz, corresponding to the first mode associated with member internal displacement and the first mode associated with nodal displacements, respectively. In both scenarios, the solutions obtained via the FEA method proved inaccurate compared to those derived from the CSD method. As seen in Figs. 9 and 9, the CSD method accurately predicts the transient response at the 50 Hz excitation frequency with only a small number of internal terms for member displacements. The results for $N_l$ = 6 and 12, and $N_t$ = 15 and 30, obtained by the CSD method show excellent agreement, demonstrating the rapid convergence of the method. Additionally, the ability to capture higher frequency vibrations further underscores the superiority of the CSD method, a capability the FEA method lacks. Similar observations are made at the 650 Hz excitation frequency, as shown in Figs. 10 and 11, which further confirms the accuracy and robustness of the CSD method.

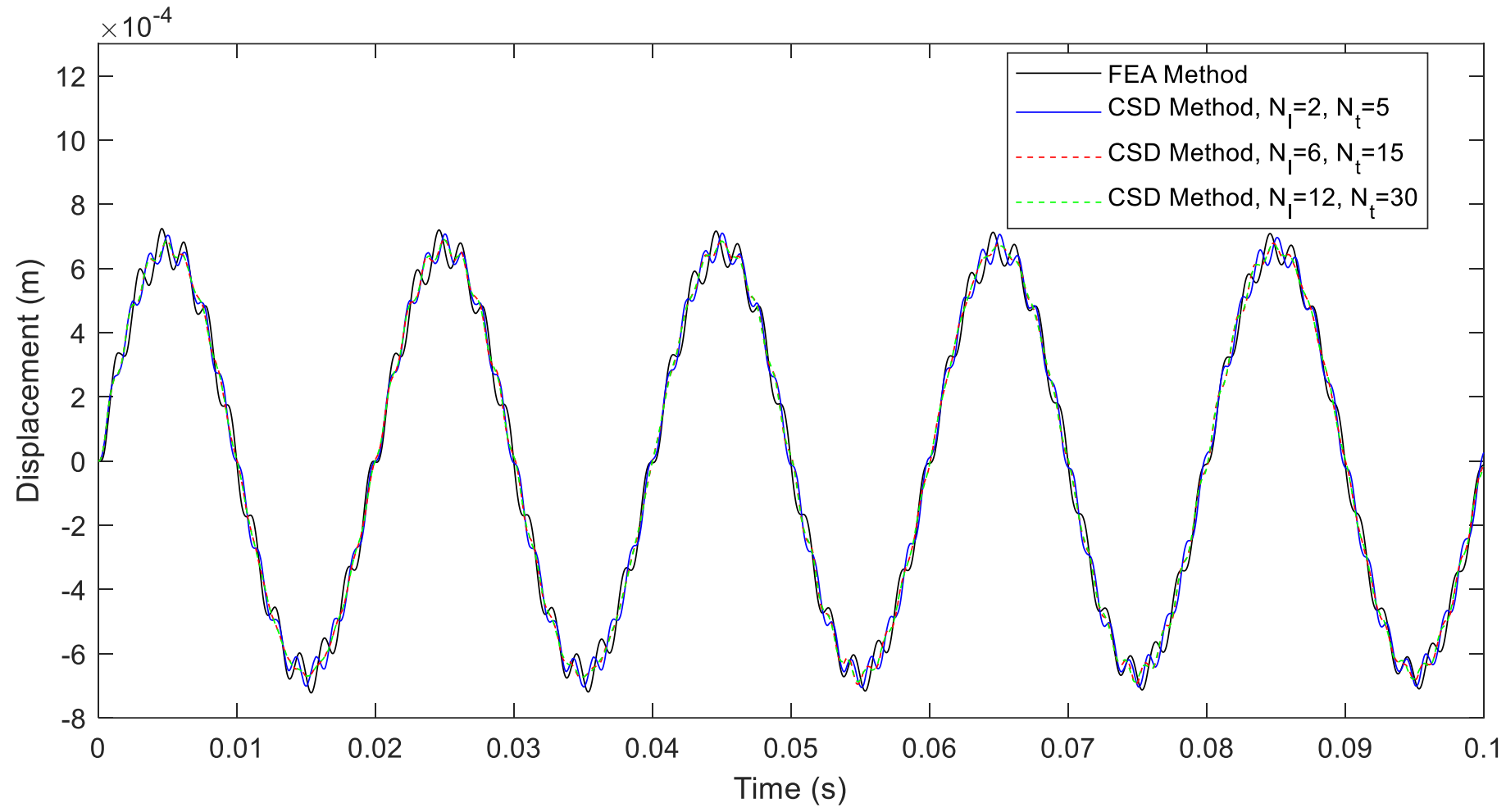

**Fig. 8  Displacement at node one of the planar cable-network structure at the excitation frequency of 50 Hz in the *x*-direction.**

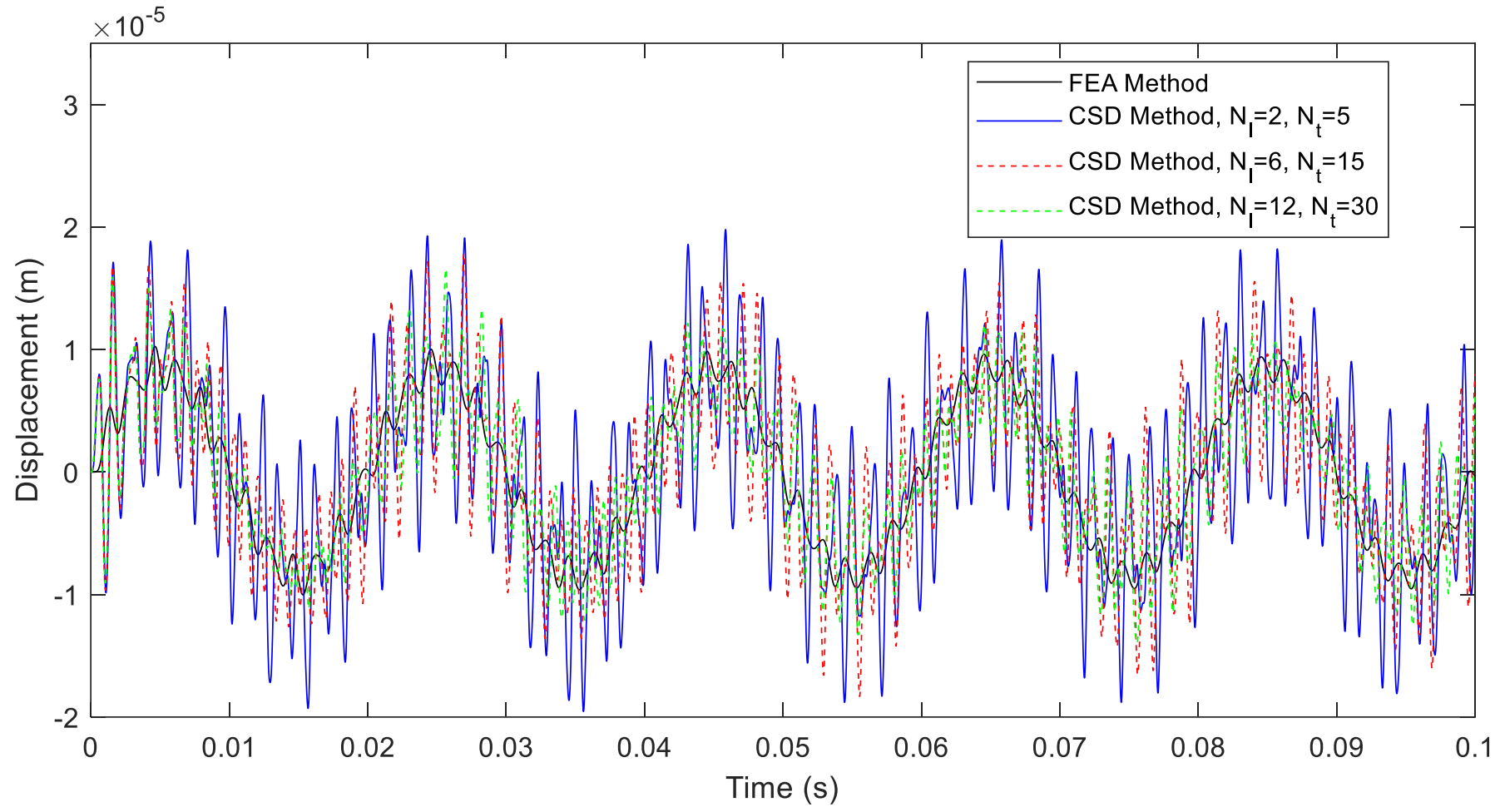

**Fig. 9  Displacement at node one of the planar cable-network structure at the excitation frequency of 50 Hz in the *y*-direction.**

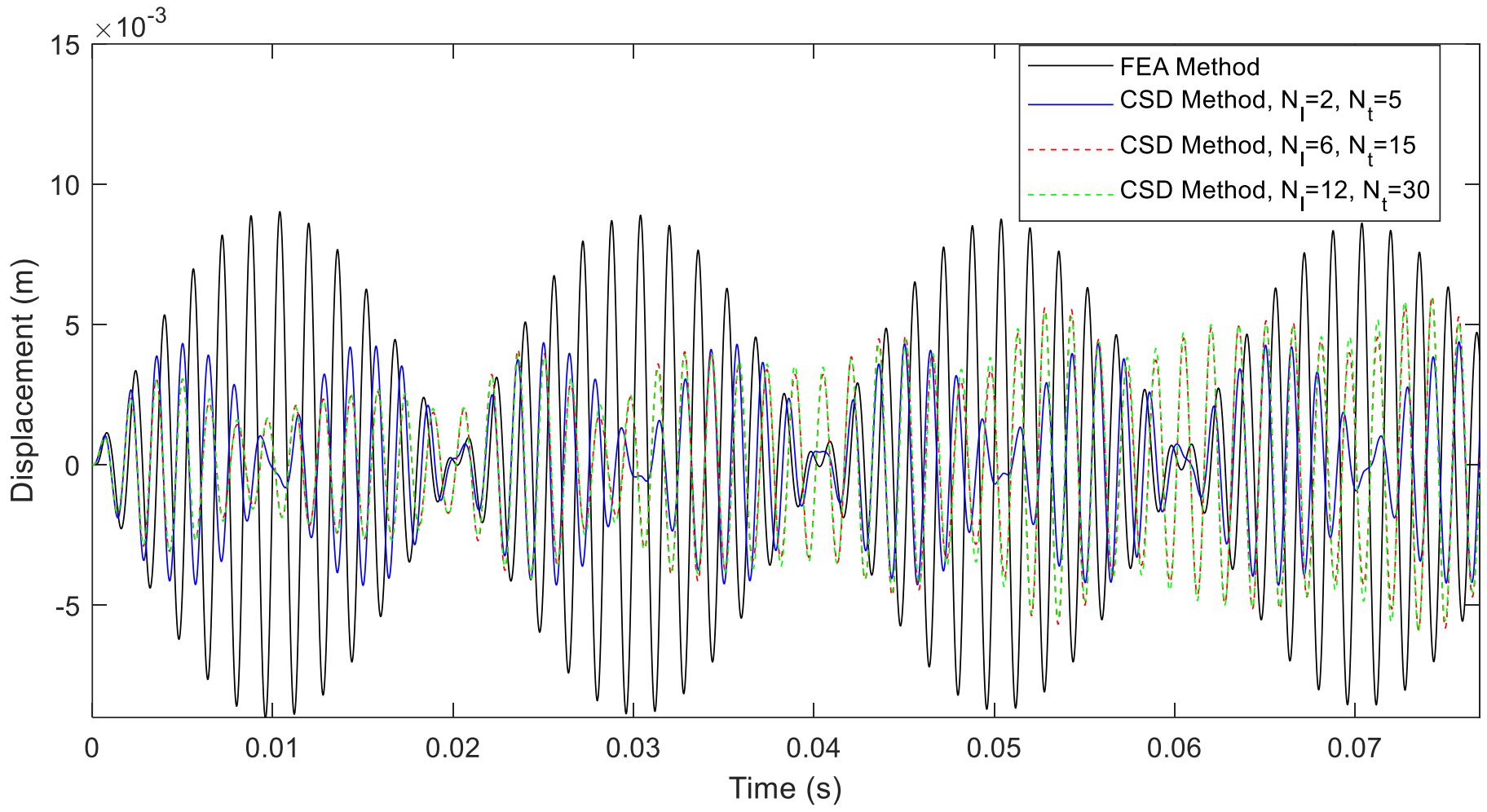

**Fig. 10 Displacement at node one of the planar cable-network structure at the excitation frequency of 650 Hz in the *x*-direction.**

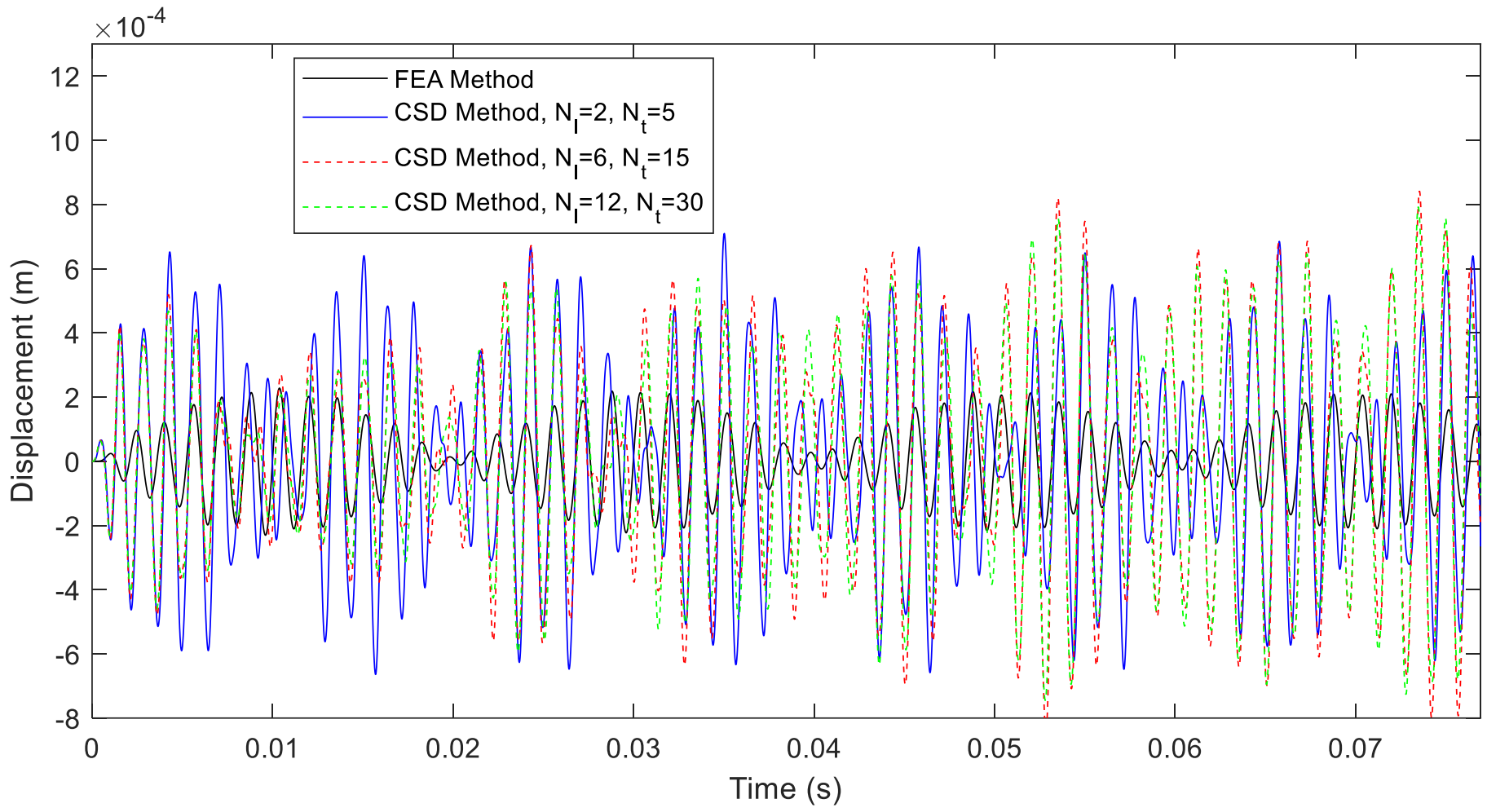

**Fig. 11 Displacement at node one of the planar cable-network structure at the excitation frequency of 650 Hz in the *y*-direction.**

### B. A center-feed parabolic DMR with 101 nodes.

In this study, the dynamic analysis of a center-feed, parabola-shaped DMR with 101 nodes is examined. The DMR features a focal length of 12 meters and an aperture diameter of 12 meters. The topology design of the cable net utilizes the technique of reducing the number of boundary nodes, as outlined in Ref [38]. The DMR structure consists of a front net, a rear net, and tension ties, with the rear net designed to mirror the front net. A three-dimensional view of the reflector is presented in Fig. 12 (a). Boundary nodes attached to the supporting ring structure are assumed to be fixed. Each cable member possesses identical material properties, characterized by a Young's modulus of 200 *GPa*, a uniform cross-sectional area of 3.14 $mm^2$, and a material density of 7850 $kg/m^3$. The tension distribution within the cable members of the front net and the tension ties is depicted in Fig. 13, with the rear net exhibiting the same tension distribution as the front net.

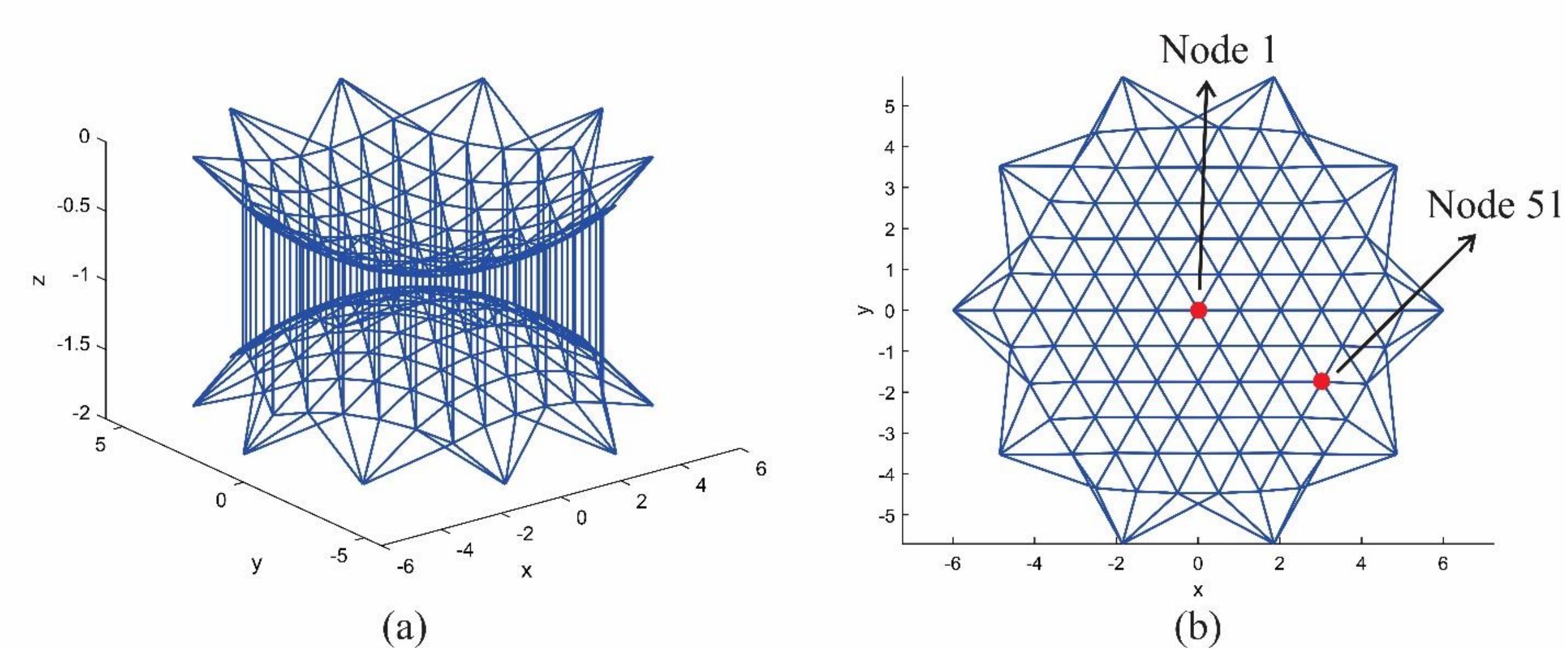

**Fig. 12 A center-feed, parabola-shaped DMR with 101 nodes: (a) perspective view; and (b) top view.**

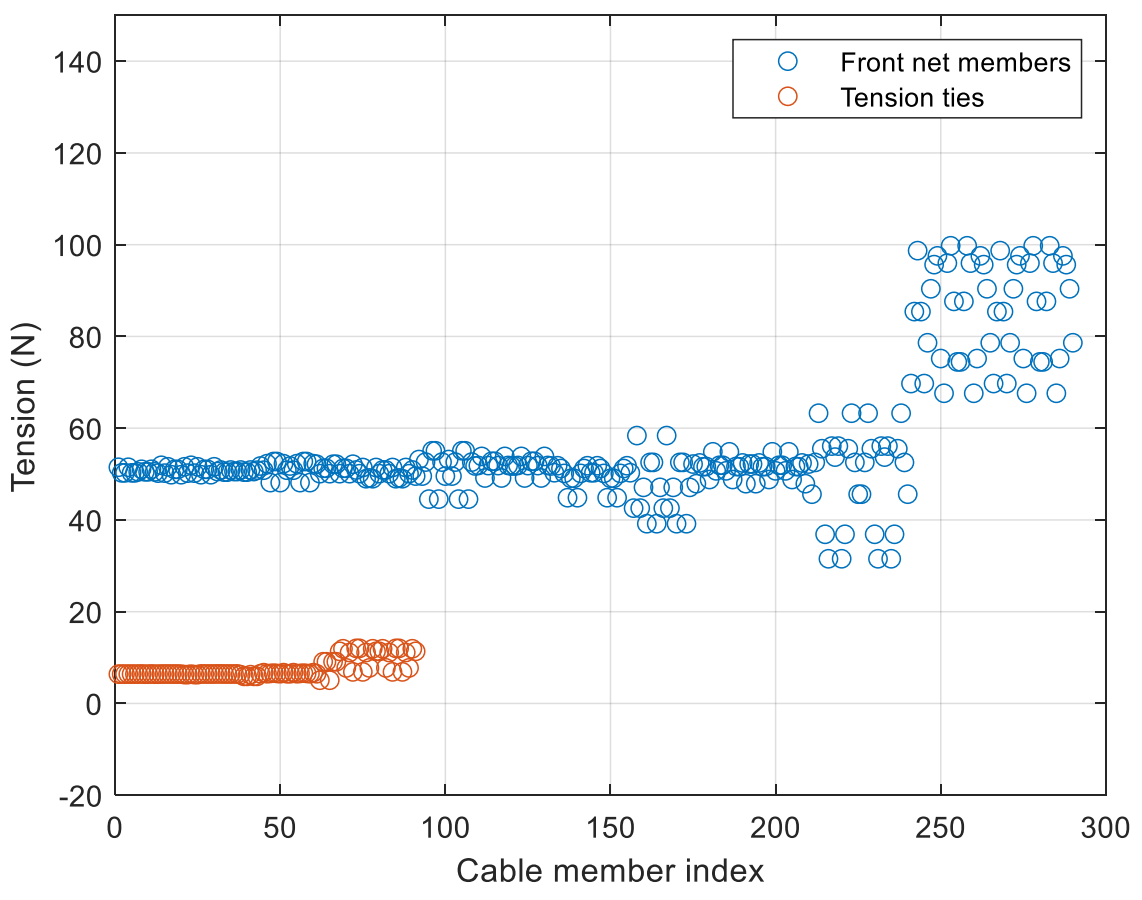

**Fig. 13 The cable member tension distribution of the 101-node DMR.**

The modal analysis of the DMR is conducted using both the FEA and CSD methods. The first 10 natural frequencies obtained from these methods are compared in Table 4. As indicated in the table, the FEA method exhibits a 2%-24% error when compared to the CSD method. Figures 14 and 15 provide a comparison of the first mode shape and the 10th mode shape of the DMR, obtained through both the FEA and CSD methods, respectively. Significant differences are observed in both mode shapes, highlighting the inaccuracies of the FEA method. In particular, the 10th mode shape obtained via the CSD method, shown in Fig. 15 (b), reveals that nodal displacements are coupled with member internal displacements. This coupling indicates the necessity of incorporating member internal displacements in dynamic modeling to achieve accurate vibration analysis.

**Table 4 The first 10 natural frequencies in Hz of the center-feed DMR obtained by the FEA and the CSD methods.**

| | $f_1$ | $f_2$ | $f_3$ | $f_4$ | $f_5$ |
|---|---|---|---|---|---|
| FEA method | 24.0585 | 24.2070 | 24.7950 | 26.9015 | 26.9793 |
| CSD method | 23.5215 | 23.6626 | 23.7627 | 23.7627 | 23.7653 |
| | $f_6$ | $f_7$ | $f_8$ | $f_9$ | $f_{10}$ |
| FEA method | 29.5038 | 29.8202 | 30.7140 | 32.3808 | 32.5604 |
| CSD method | 23.7654 | 24.2438 | 26.1163 | 26.1660 | 28.3935 |

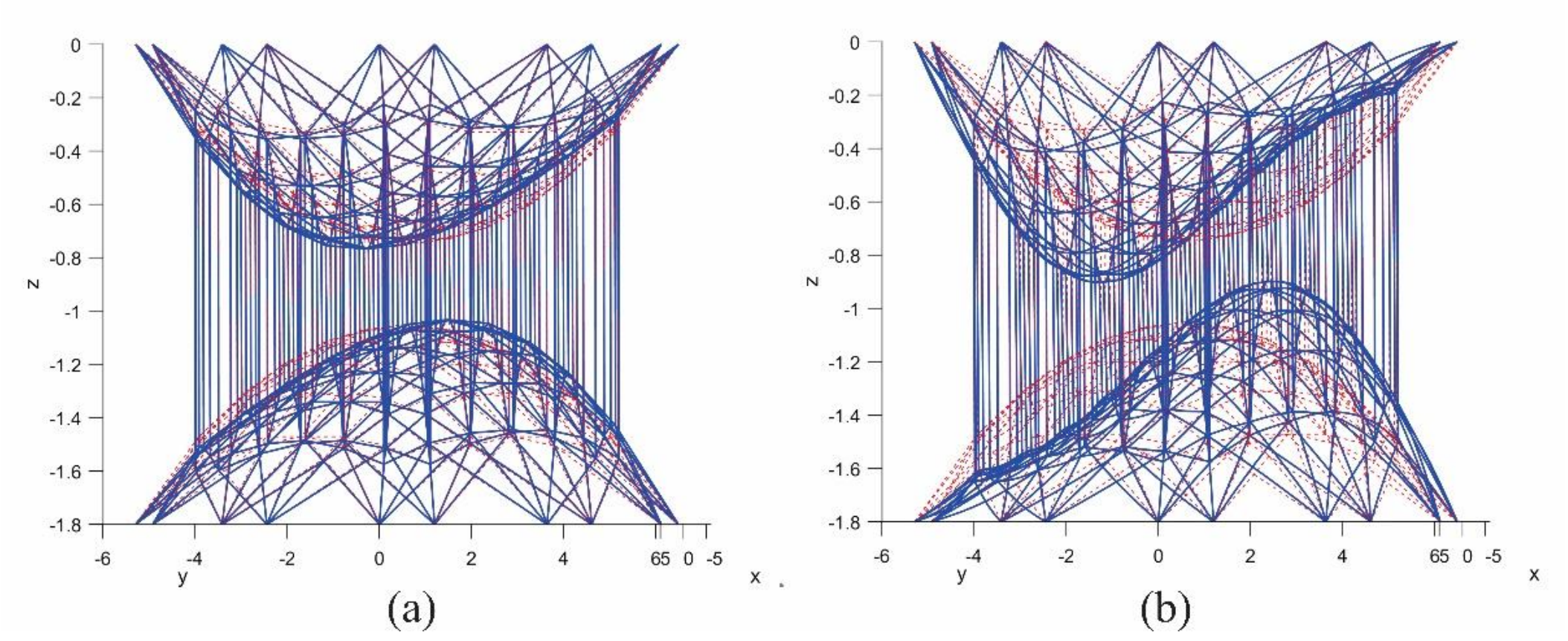

**Fig. 14 The first mode shapes of the center-feed DMR obtained by (a) the FEA method and (b) the CSD method.**

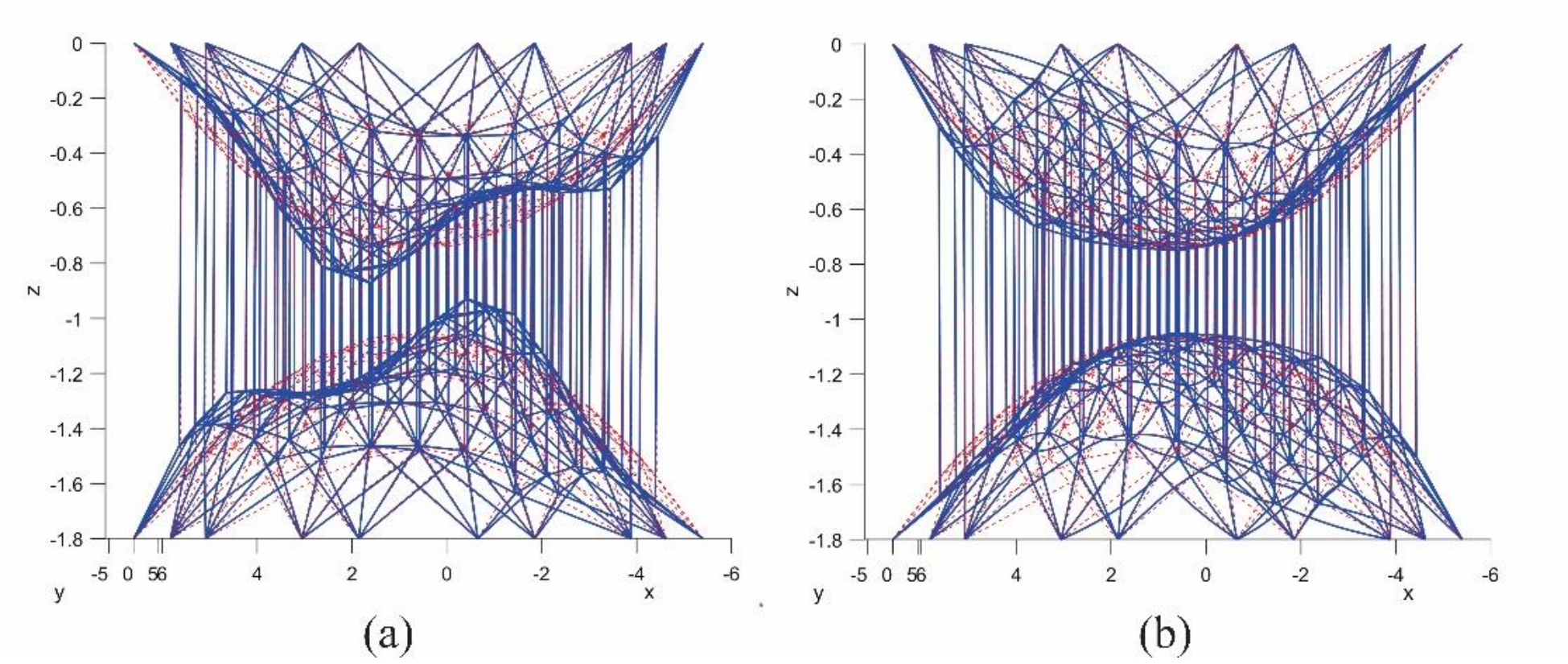

**Fig. 15 The 10th mode shapes of the center-feed DMR obtained by (a) the FEA method and (b) the CSD method.**

The frequency response analysis of the DMR is performed by applying a point-wise sinusoidal force $F_f$ = $F_0\sin(2\pi ft)$ at node one in the *z*-direction, where $F_0$ = 100*N*. The frequency response of node 51 in the *z*-direction, within the 10-50 Hz range, is predicted using both the FEA and CSD methods. The locations of nodes 1 and 51 are illustrated in Fig. 12 (b). The results, given in Fig. 16, obtained from the FEA and CSD methods show good agreement under 20 Hz, a frequency near the first natural frequency, thereby validating the correctness of the CSD method. However, a noticeable deviation between the results of the two methods occurs for frequencies above 25 Hz, highlighting the inaccuracy of the FEA method for higher frequency range. Additionally, the convergence of the CSD method is evident with the increasing number of internal terms used in dynamic modeling.

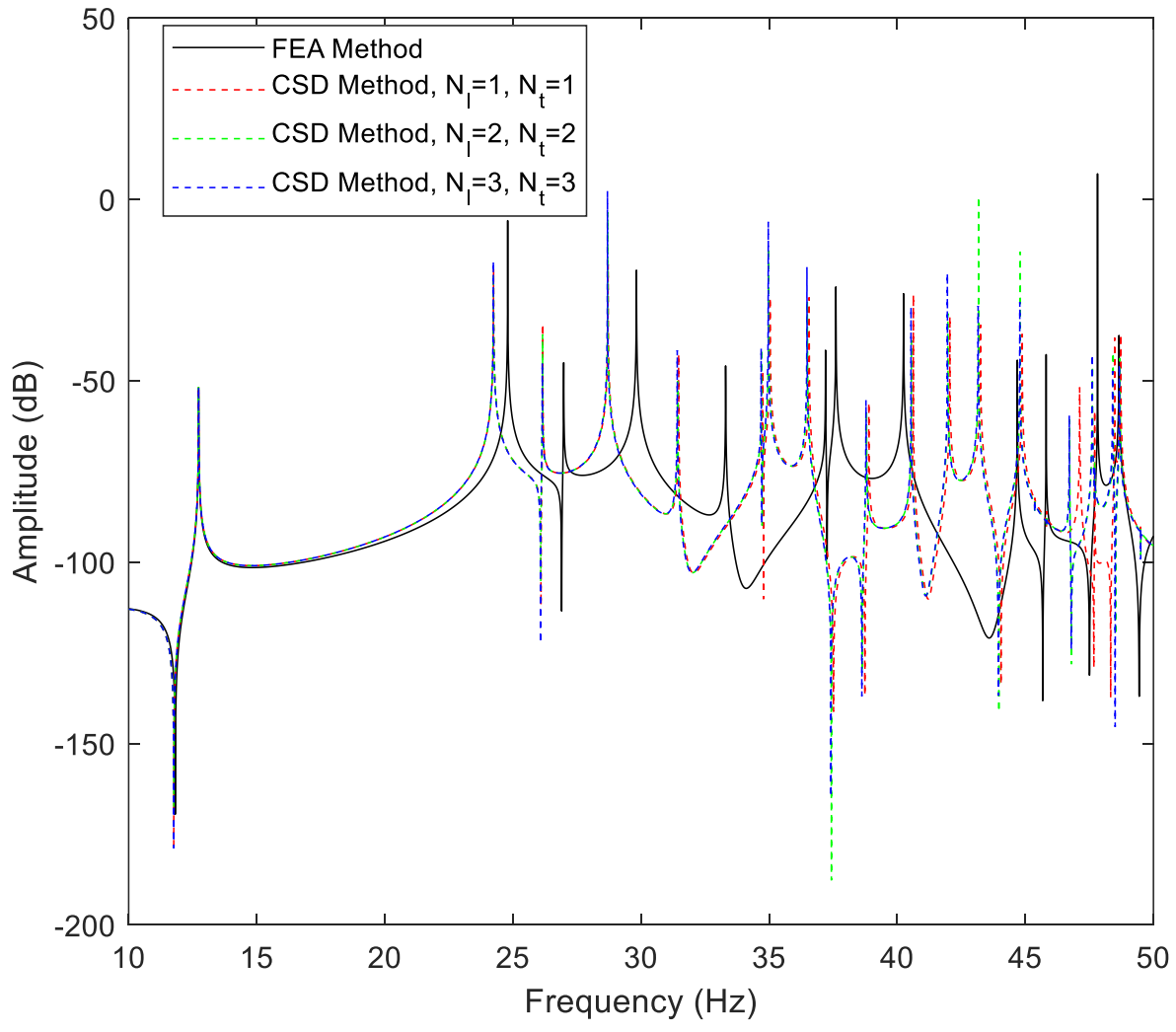

**Fig. 16  Frequency responses at node 51 of the 101-node DMR at 10-50 Hz obtained by the FEA method and the CSD method.**

The transient response analysis of the DMR is conducted by applying a point-wise sinusoidal force $F_f = F_0\sin(2\pi ft)$ at node one in the $z$-direction, where $F_0 = 100N$. The transient responses of node 51 in the $z$-direction are plotted in Figs. 17 and 18 at excitation frequencies of 20 Hz and 50 Hz, respectively. Consistent with the results observed for the planar cable-network structure, the solutions obtained via the FEA method proved inaccurate compared to those derived from the CSD method in both scenarios. The CSD method accurately predicts the transient response at both the 20 Hz and 50 Hz excitation frequencies, even with a small number of internal terms of member displacements being used.

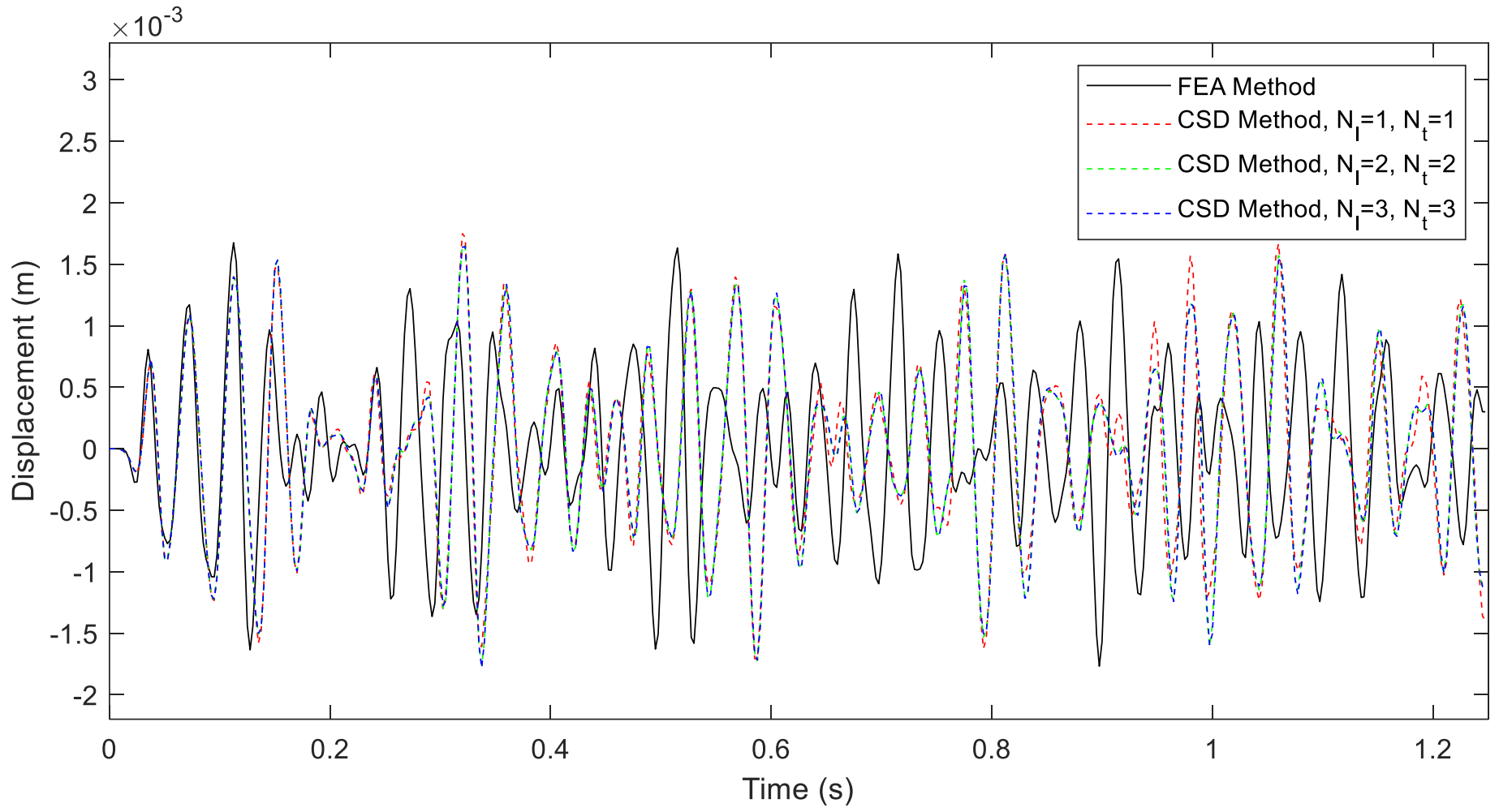

**Fig. 17  Displacement in the *z*-direction at node 51 of the 101-node DMR at the excitation frequency of 20 Hz.**

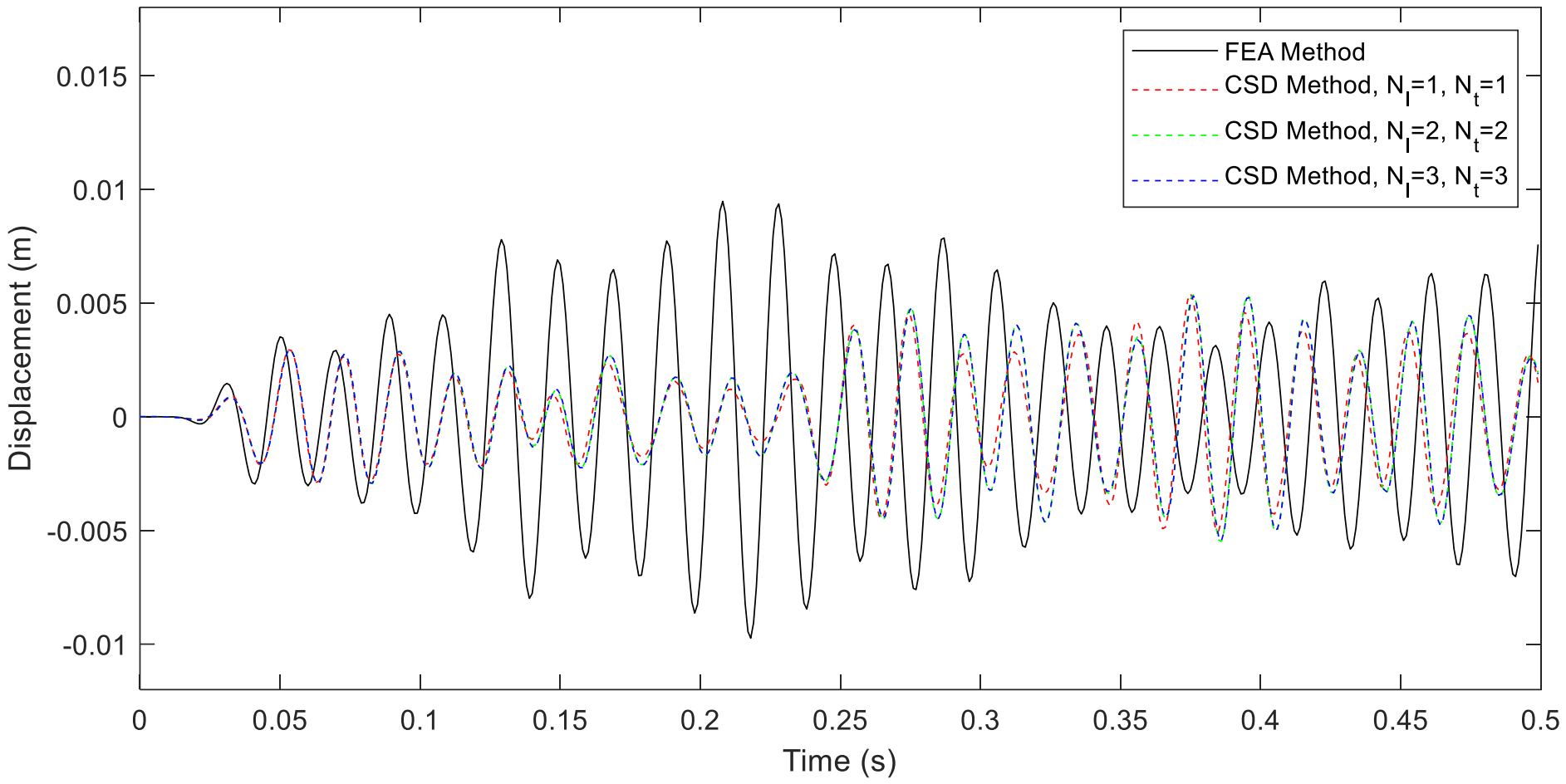

**Fig. 18 Displacement in the *z*-direction at node 51 of the 101-node DMR at the excitation frequency of 50 Hz.**

## VI.Conclusion

A novel method, named the Cartesian spatial discretization method, for dynamic modeling and vibration analysis of large deployable mesh reflectors is proposed in this paper. This method models the positions of cable members as a summation of internal and boundary-induced terms within a global Cartesian coordinate system, deriving a nonlinear dynamic model from Lagrange's equations. For vibration analysis, the model is linearized around the equilibrium configuration of the reflector and assembled by integrating common nodal coordinates of individual cable members. Unlike traditional wave approaches that are confined to planar cable nets, this method accommodates three-dimensional structures through the use of a global Cartesian coordinate system, ensuring accurate dynamic response predictions for both simple and complex deployable mesh reflector configurations. Another significant advantage of the proposed method is its comprehensive integration of member internal displacements within the dynamic model, thereby avoiding the oversimplification of cable members prevalent in conventional finite element approaches. Simulation results confirm the effectiveness of the proposed method over traditional dynamic modeling methods, highlighting its superiority in capturing the complex dynamic behavior of deployable mesh reflectors. The comprehensive analyses presented demonstrate that the proposed method consistently provides accurate and reliable predictions, thereby validating its applicability and robustness in practical engineering applications.

# Appendix

Consider a two-dimensional cable member shown in Fig. 19 with one internal term in the longitudinal direction and one internal term in the transverse direction. Coordinates of the two nodes of the member are ($x_0$, $y_0$) and ($x_1$, $y_1$). The longitudinal direction $\vec{r}$ and transverse direction $\vec{w}$ are obtained as

$$\vec{r} = \begin{bmatrix} (x_1 - x_0)/L \\ (y_1 - y_0)/L \end{bmatrix}, \quad \vec{w} = \begin{bmatrix} (y_0 - y_1)/L \\ (x_1 - x_0)/L \end{bmatrix} \tag{A.1}$$

where $L = \sqrt{(x_1 - x_0)^2 + (y_1 - y_0)^2}$ . Position $u$ in Eq. (13) is expressed as

$$u = \begin{bmatrix} x_1\xi - x_0(\xi - 1) - \dfrac{x_0 - x_1}{L}\sin(\pi\xi)q_l + \dfrac{y_0 - y_1}{L}\sin(\pi\xi)q_t \\ y_1\xi - y_0(\xi - 1) - \dfrac{y_0 - y_1}{L}\sin(\pi\xi)q_l - \dfrac{x_0 - x_1}{L}\sin(\pi\xi)q_t \end{bmatrix} \tag{A.2}$$

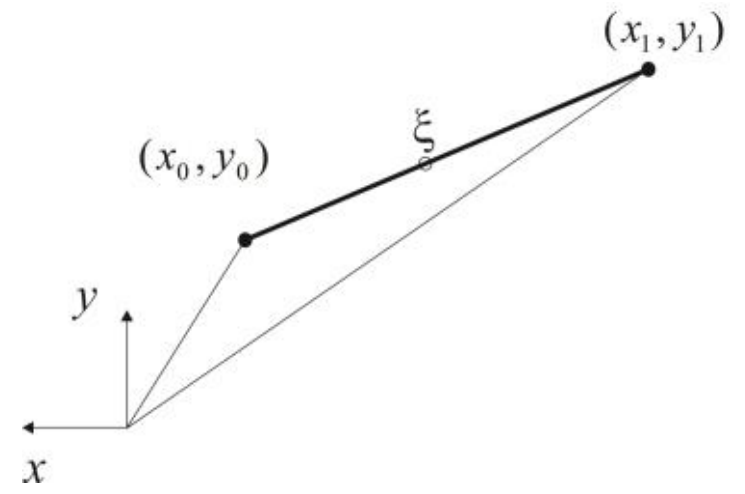

**Fig. 19 Motion of a two-dimensional cable member.**

Velocity $\dot{u}$ in Eq. (19) can be written as

$$\dot{u} = \begin{bmatrix} \dot{x}_1\xi - \dot{x}_0(\xi - 1) - \dfrac{x_0 - x_1}{L}\sin(\pi\xi)\dot{q}_l + \dfrac{y_0 - y_1}{L}\sin(\pi\xi)\dot{q}_t \\ \dot{y}_1\xi - \dot{y}_0(\xi - 1) - \dfrac{y_0 - y_1}{L}\sin(\pi\xi)\dot{q}_l - \dfrac{x_0 - x_1}{L}\sin(\pi\xi)\dot{q}_t \end{bmatrix} \tag{A.3}$$

The kinetic energy of the cable member in Eq. (21) is obtained as

$$\begin{aligned} T = \frac{m}{24L^2\pi^2}\Big[& 4L^2\pi^2(\dot{x}_0^2 + \dot{x}_1^2 + \dot{y}_0^2 + \dot{y}_1^2) + 6\pi^2(\dot{q}_l^2 + \dot{q}_t^2)(x_0^2 + x_1^2 + y_0^2 + y_1^2) \\ & + 4L^2\pi^2(\dot{x}_0\dot{x}_1 + \dot{y}_0\dot{y}_1) - 12\pi^2(\dot{q}_l^2 + \dot{q}_t^2)(x_0x_1 + y_0y_1) \\ & - 24L\pi(\dot{q}_l\dot{x}_0x_0 - \dot{q}_l\dot{x}_0x_1 + \dot{q}_t\dot{y}_0x_0 - \dot{q}_t\dot{y}_0x_1) \\ & + 24L\pi(\dot{q}_l\dot{y}_0y_0 - \dot{q}_l\dot{y}_0y_1 + \dot{q}_t\dot{x}_0y_0 - \dot{q}_t\dot{x}_0y_1)\Big] \end{aligned} \tag{A.4}$$

The potential energy of the cable member in Eq. (28) is obtained as

$$V = EA\left[\frac{(L - L_0)^2}{2L_0} + \frac{\pi^2 q_l^2}{4L_0} - \frac{\pi q_t^2}{4L} + \frac{\pi^2 q_t^2}{4L_0}\right] \tag{A.5}$$

The generalized coordinate of the cable member is defined as

$$q = \begin{bmatrix} x_0 & y_0 & x_1 & y_1 & q_l & q_t \end{bmatrix}^T \tag{A.6}$$

the linear mass and stiffness matrices in Eq. (31) are expressed as

$$M = m\begin{bmatrix} \frac{1}{3} & 0 & \frac{1}{6} & 0 & \frac{x_1 - x_0}{\pi L} & \frac{y_0 - y_1}{\pi L} \\ 0 & \frac{1}{3} & 0 & \frac{1}{6} & \frac{y_1 - y_0}{\pi L} & \frac{x_0 - x_1}{\pi L} \\ \frac{1}{6} & 0 & \frac{1}{3} & 0 & \frac{x_1 - x_0}{\pi L} & \frac{y_0 - y_1}{\pi L} \\ 0 & \frac{1}{6} & 0 & \frac{1}{3} & \frac{y_1 - y_0}{\pi L} & \frac{x_1 - x_0}{\pi L} \\ \frac{x_1 - x_0}{\pi L} & \frac{y_1 - y_0}{\pi L} & \frac{x_1 - x_0}{\pi L} & \frac{y_1 - y_0}{\pi L} & \frac{1}{2} & 0 \\ \frac{y_0 - y_1}{\pi L} & \frac{x_0 - x_1}{\pi L} & \frac{y_0 - y_1}{\pi L} & \frac{x_1 - x_0}{\pi L} & 0 & \frac{1}{2} \end{bmatrix}, \quad K = \begin{bmatrix} k_{11} & k_{12} & k_{13} & k_{14} & 0 & 0 \\ k_{12} & k_{22} & k_{23} & k_{24} & 0 & 0 \\ k_{13} & k_{23} & k_{33} & k_{34} & 0 & 0 \\ k_{14} & k_{24} & k_{34} & k_{44} & 0 & 0 \\ 0 & 0 & 0 & 0 & k_{55} & 0 \\ 0 & 0 & 0 & 0 & 0 & k_{66} \end{bmatrix} \tag{A.7}$$

where

$$
\begin{aligned}
k_{11} &= \frac{EA(2x_0 - 2x_1)^2}{4L_0 L^2} - \frac{EA(2L_0 - 2L)(y_0 - y_1)^2}{2L_0 L^3}, \\
k_{12} &= \frac{EA(2x_0 - 2x_1)(2y_0 - 2y_1)}{4L_0 L^2} + \frac{EA(2L_0 - 2L)(x_0 - x_1)(y_0 - y_1)}{2L_0 L^3}, \\
k_{13} &= \frac{EA(2L_0 - 2L)(y_0 - y_1)^2}{2L_0 L^3} - \frac{EA(2x_0 - 2x_1)^2}{4L_0 L^2}, \\
k_{14} &= -\frac{EA(2x_0 - 2x_1)(2y_0 - 2y_1)}{4L_0 L^2} - \frac{EA(2L_0 - 2L)(x_0 - x_1)(y_0 - y_1)}{2L_0 L^3}, \\
k_{22} &= \frac{EA(2y_0 - 2y_1)^2}{4L_0 L^2} - \frac{EA(2L_0 - 2L)(x_0 - x_1)^2}{2L_0 L^3}, \\
k_{23} &= -\frac{EA(2x_0 - 2x_1)(2y_0 - 2y_1)}{4L_0 L^2} - \frac{EA(2L_0 - 2L)(x_0 - x_1)(y_0 - y_1)}{2L_0 L^3}, \\
k_{24} &= \frac{EA(2L_0 - 2L)(x_0 - x_1)^2}{2L_0 L^3} - \frac{EA(2y_0 - 2y_1)^2}{4L_0 L^2}, \\
k_{33} &= \frac{EA(2x_0 - 2x_1)^2}{4L_0 L^2} - \frac{EA(2L_0 - 2L)(y_0 - y_1)^2}{2L_0 L^3}, \\
k_{34} &= \frac{EA(2x_0 - 2x_1)(2y_0 - 2y_1)}{4L_0 L^2} + \frac{EA(2L_0 - 2L)(x_0 - x_1)(y_0 - y_1)}{2L_0 L^3}, \\
k_{44} &= \frac{EA(2y_0 - 2y_1)^2}{4L_0 L^2} - \frac{EA(2L_0 - 2L)(x_0 - x_1)^2}{2L_0 L^3}, \\
k_{55} &= \frac{EA\pi^2}{2L_0}, \\
k_{66} &= \frac{EA\pi^2}{2L_0} - \frac{EA\pi^2}{2L}
\end{aligned}
\tag{A.8}
$$

## Funding Sources

This research was supported by the US National Science Foundation through grant number 2335692.